\documentclass[%
 reprint,
 amsmath,amssymb,
 aps,
]{revtex4-2}
\usepackage{mathtools, nccmath}
\DeclarePairedDelimiter{\nint}\lfloor\rceil
\usepackage{graphicx}
\usepackage{dcolumn}
\usepackage{bm}
\usepackage{tabularx,booktabs}
\usepackage{hyperref}
\usepackage{xcolor}
\usepackage{multirow}

\preprint{APS/123-QED}
\begin{document}
\title{\textcolor{blue}{Feature Learning} and \textcolor{blue}{Network Structure} from Noisy \textcolor{blue}{Node Activity} Data}

\author{Junyao Kuang$^1$}
\email{kuang@ksu.edu}
\author{Caterina Scoglio$^1$}
\author{Kristin Michel$^2$}

\affiliation{$^1$Department of Electrical and Computer Engineering}
\affiliation{$^2$Division of Biology, Kansas State University}

\date{\today}

\begin{abstract}
In the \textcolor{blue}{studies} of network structures, much attention has been devoted to \textcolor{blue}{developing approaches to reconstruct networks and predict missing links when edge-related information is given}. However, \textcolor{blue}{such approaches are not applicable when we are only given noisy node activity data} with missing values. \textcolor{blue}{This work presents} an unsupervised learning framework to learn node vectors and construct networks from such \textcolor{blue}{node activity data}. \textcolor{blue}{First, we design a scheme to generate random node sequences from node context sets, which are generated from node activity data.} Then, a three-layer neural network is adopted training the node sequences to obtain node vectors, which \textcolor{blue}{allow us to construct networks and} capture nodes with synergistic roles. Furthermore, we present an entropy-based approach to select the most meaningful neighbors for each node in the \textcolor{blue}{resulting network}. Finally, the effectiveness of the method is validated through both synthetic and real data.

\end{abstract}

\maketitle
\section{Introduction}
A network is a system-level view of pairwise interactions between nodes, genes, \textcolor{blue}{or} elements in a complex system \cite{barabasi,barabasi1,bacco,kuang,newman1,newman3,decelle, pardo,peixoto1,peixoto2,van1,peixoto3,newman4,peixoto4}. The first step in analyzing a networked system is to construct the network from data obtained with different technologies. \textcolor{blue}{In most cases,} network structures \textcolor{blue}{can be determined} through direct measurements, \textcolor{blue}{meaning that pairwise relationships between nodes can be} observed directly. For instance, the edges in friendship networks can be probed through various ways, including using questionnaires, checking Facebook or Twitter friendship, and investigating face-to-face interactions \cite{timme1, timme2, van2, bianconi, newman2}. As another example, edges in web graphs can be directly determined by checking if hyperlinks exist between web pages. \textcolor{blue}{However, there are cases where the relationships between nodes cannot be observed directly \cite{bob}.} \textcolor{blue}{Instead, we may only have node activity data that reflect the properties of nodes from various aspects. In these cases,} we need to estimate the underlying network structure \textcolor{blue}{from nodal data}. \textcolor{blue}{Such problem exists in many areas, including} the construction of financial, biological, and climate networks \cite{horvath, lynall, raimondo, eco, neuro, neuro1, neuro2, mibrain, fin, climate, climate2}. In these areas, \textcolor{blue}{measurements of pairwise relationships are not always feasible} \cite{bob, newmannc, newmanpre1}. Instead, \textcolor{blue}{we can conduct various experiments to measure node activities under different conditions} \cite{barabasi2}.

This work develops a model to learn node representations from noisy and heterogeneous data and proposes an entropy-based method to extract network structures. Specifically, we investigate the problems of feature learning and network construction for gene co-expression data. \textcolor{blue}{Different} high throughput technologies, including \textcolor{blue}{microarray} and RNA-sequencing, allow simultaneously evaluating thousands of gene expression data. Usually, the data can be organized into a matrix that consists of rows representing $N$ genes (nodes) and columns representing $M$ experimental conditions. To construct a network from such expression data, we need to consider three problems. (1) The expression data, \textcolor{blue}{measured} through different experimental technologies, are distributed in various ranges. For example, the raw expression values obtained from different versions of RNA-sequencing in different labs are dispersed from zero to tens of thousands and do not follow any specific distribution. (2) Missing values are frequently present in the dataset. Some experiments may only test a subset of genes for specific purposes, or some experimental data for some genes (nodes) are not available. (3) The levels of noise are not constant. For instance, the environments, such as humidity, temperature, and light intensity, could potentially influence the accuracy of the devices and the measured node activity data. \textcolor{blue}{The method for network construction is not allowed to be affected by missing values and noisy data.}

There are diverse approaches aiming at constructing networks from nodal data. A significant volume of works uses the correlation coefficient to measure the degree to which a pair of nodes is related, and edges are selected by thresholding the correlation coefficients \cite{horvath1, horvath2, horvath3}. However, \textcolor{blue}{the drawbacks of the correlation methods are that: 1) the expression data are required to follow a (quasi-) normal distribution, 2) the correlation coefficients are significantly affected by outliers, and 3) the number of measured conditions and missing values substantially affect the results.} \cite{horvath1, horvath2}. Mutual Information (MI) and its variants are also used to construct gene co-expression networks. The MI models do not require the data to follow the normal distribution. Still, the MI models are even more complex, since we are expected to find the joint probability distribution for every pair of genes \cite{info}. We need to solve the problems mentioned above before applying either of the two methods. To solve problems (1) and (3), \textcolor{blue}{some researchers have proposed using rescaling and normalization methodologies to obtain quasi-normal distributed data from the raw node activity data} \cite{cancer}. According to \cite{reverter}, the number of experimental conditions significantly influences the correlation coefficients under the null hypothesis that two nodes are not correlated. Theoretical analysis shows that correlations based on ten conditions tend to be higher than those computed with 50 conditions. Missing values lead to node pairs with a different number of paired elements, meaning that the node pairs with fewer paired elements are more likely to have high correlation coefficients. \textcolor{blue}{Therefore, some works use imputation or interpolation to solve problem (2) \cite{misdata, kuangano}. The complex data processing procedures pose a severe challenge for the principle of parsimony when we further study the resulting network structure \cite{occam}.}   

Edge selection is another issue we need to consider when \textcolor{blue}{constructing networks from node activity data}. Both correlation and MI methods return coefficients between -1 and 1. \textcolor{blue}{Many researchers construct} unweighted networks by applying a threshold to select \textcolor{blue}{edges of the network corresponding to node pairs with the highest coefficients}. However, choosing a threshold is always tricky since a high threshold could generate singleton nodes, while a low threshold generates networks with many weakly connected node pairs \cite{cancer, dam}. \textcolor{blue}{Though the problem can be solved by fixing the minimum number of neighbors of each node, the choice of the threshold influences the node degree distribution, meaning that nodes' roles in the resulting network are related to the choice of thresholds. As an alternative, we propose an entropy-based network construction method, which has better performance in maintaining nodes' roles (e.g. hubs and leaf nodes) and avoiding isolating nodes.}

This paper proposes a neural network-based method to extract node representations, and presents an entropy-based approach to construct networks from noisy node activity data. Inspired by the application of neural networks in natural language processing (NLP) \cite{vec1,vec2,vec3,lan1,lan2,lan3}, we \textcolor{blue}{propose generating} node sequences from node activity data to simulate sentences in documents. The neural network model can embed node sequences into vectors of identical dimensions, which \textcolor{blue}{allow us to study node features and construct networks. The main contributions of the paper are as follows: First, we design a simple and direct data processing scheme to generate random node sequences from $M$ conditions. In our approach, the raw data are not required to follow any specific distribution. Thus, re-scaling and normalization are obsolete. In addition, the $M$ conditions are processed separately, meaning that negative impacts from missing data and outliers can be minimized. Second, the node sequences are trained with a three-layer neural network model, which builds on the hypothesis that nodes with similar properties tend to have similar neighbors \cite{lan3}. As a result, similar nodes have similar values in the trained node vectors. Third, we propose an entropy-based method to extract the corresponding network where selected edges can recover node roles \cite{lee,entropy2, lee}. Finally, we demonstrate the validity of the proposed approach experimentally using synthetic and real data.}

\section{Approach} \label{II} 
In this section, we define the context set, node sequence generation, and the entropy-based method for \textcolor{blue}{network construction.}

\textcolor{blue}{In human language, words} in similar contexts tend to have similar meanings \cite{vec1}. That is, words with similar meanings usually show in similar neighborhoods. We can use NLP models to learn node representations if we have node sequences in which nodes with similar measurements are in similar contexts. \textcolor{blue}{The measurements of nodes in different conditions represent different properties,} similar to words in various topics that may have different meanings. Building on these observations, we design a scalable \textcolor{blue}{node sequence} generation strategy to process the $M$ conditions separately.

\subsection{\textcolor{blue}{Generate context sets from node activity data}}\label{IIA} 
Suppose the $N$ nodes are measured in $M$ conditions. Given a node $v_i$ ($i \leq N$), we assume its value in the $\omega$th condition is $v_i(\omega)$. \textcolor{blue}{We define the} context set of node $v_i$ in the $\omega$th condition \textcolor{blue}{as:}
\begin{align}
\label{eq:s1}
C_\omega(v_i)=\{v_j :\ |v_j(\omega)-v_i(\omega)|\leq \delta_i^\omega\}.
\end{align}
where the tolerance $\delta_i^\omega$ can be a \textcolor{blue}{parameter} such that $\delta_i^\omega=\beta_\omega v_i(\omega)$. By employing the parameter $\beta_\omega$, we can tune the size of the context set per the error levels of different technologies. \textcolor{blue}{In this work, we skip the generation of context set $C_\omega(v_i)$ when a missing value is present in the $\omega$th condition for node $v_i$, and we do not predict the missing values from other conditions.}

Formally, the context set of node $v_i$ is composed of nodes with measurements falling in the range $[v_i(\omega)-\delta_i^\omega,\  v_i(\omega)+\delta_i^\omega]$. Therefore, the number of elements of the intersection set $C_\omega (v_i)\cap C_\omega (v_k)$ is related to the \textcolor{blue}{measurements} of the two nodes $v_i$ and $v_k$. For example, assume the measurements of the three nodes $v_x$, $v_y$, and $v_z$ in the $\omega$th condition are respectively $v_x (\omega)=1000$, $v_y (\omega)=990$, and $v_z (\omega)=950$. It is clear that $v_y (\omega)$ is closer to $v_x (\omega)$ than $v_z (\omega)$, i.e., $|v_x(\omega)-v_y(\omega)|<|v_x(\omega)-v_z(\omega)|$. Therefore, we have the following inequality:
\begin{align}
\label{eq:s2}
|C_\omega (v_x)\cap C_\omega (v_y)|\geq |C_\omega (v_x)\cap C_\omega (v_z)|,
\end{align}
where $|\cdot|$ denotes the cardinality of the intersection set. The context set $C_\omega (v_y )$ recapitulates more elements of $C_\omega (v_x)$ than the context set $C_\omega (v_z )$. For any node $v_j \in C_\omega (v_x)$, we have the probability
\begin{align}
\label{eq:s3}
P(v_j \in C_\omega (v_y)) \geq P(v_j \in C_\omega (v_z)).
\end{align}

Furthermore, we assume $G_\omega (v_i)$ is a set consisting of nodes whose context set contains node $v_i$, such that
\begin{align}
\label{eq:s4}
G_\omega (v_i)=\{ v_j :\ v_i\in C_\omega (v_j)\}.
\end{align}

Based on the same example above, we can say that there are more context sets containing simultaneously $v_x$ and $v_y$ than containing simultaneously $v_x$ and $v_z$. Therefore, we have 
\begin{align}
\label{eq:s5}
|G_\omega (v_x) \cap G_\omega (v_y)| \geq |G_\omega (v_x) \cap G_\omega (v_z)|,
\end{align}
meaning that the nodes with closer values are more likely to be present in the same context sets. Similarly, for any node $v_j \in G_\omega (v_x)$, we have the probability
\begin{align}
\label{eq:s6}
P(v_j \in G_\omega (v_y)) \geq P(v_j \in G_\omega (v_z)).
\end{align}

In the generation of node sequences, we always sample the subsequent node from the context set of the current node. For example, given a node sequence $l$, suppose the $i$th node is $v_x$, i.e., $l_i=v_x$. Then, \textcolor{blue}{we have a node sequence}
\begin{align}
\label{eq:s7}
\{\cdots,\ l_{i-1}\in G_\omega(v_x),\ l_i=v_x,\ l_{i+1}\in C_\omega(v_x),\ \cdots\}.
\end{align}

Based on Eq. \ref{eq:s3} and Eq. \ref{eq:s6}, $l_{i+1}$ tends to be in $C_\omega(v_y)$ \textcolor{blue}{with higher probability} than $C_\omega(v_z)$, and $l_{i-1}$ is more likely to be in $G_\omega(v_y)$ than \textcolor{blue}{in} $G_\omega(v_z)$. That is, the context nodes of $v_x$ tend to be the context nodes of $v_y$ rather than $v_z$, since $v_y(\omega)$ is closer to $v_x(\omega)$ than $v_z(\omega)$. Therefore, in the generated node sequences, we can say that nodes with closer values tend to appear in similar contexts.

\subsection{\textcolor{blue}{Generate random node sequences}}\label{IIB} 
The simplest way to generate node sequences \textcolor{blue}{from context sets} would be to \textcolor{blue}{randomly} sample the next node from the context set of the current node, which is exactly the first order Markov chain \cite{mark}. Assume the $i$th node of a node sequence is $l_i$, the next node $l_{i+1} \in C_\omega(l_i)$ is chosen with probability
\begin{align}
\label{eq:s8}
p(l_{i+1}\ |\ l_{i})=\frac{1}{|C_\omega(l_i)|}.
\end{align}

Under this assumption, the nodes in the context set $C_\omega(l_i)$ have \textcolor{blue}{an equal probability of being} chosen as the subsequent node.

Alternatively, we can generate biased random node sequences. Suppose we have just traversed node $l_{i-1}$, and now we reside at node $l_i$. The probability of sampling the next node $l_{i+1}$ is biased by the previous node $l_{i-1}$. Therefore, we introduce a parameter $\rho$, and the unnormalized probability of the next node is\textcolor{blue}{
\begin{align}
\label{eq:s9}
p(l_{i+1}\ |\ l_{i}, l_{i-1})=
\begin{cases}
      1 & \text{if if $l_{i+1} \in C_\omega(l_{i-1}) \cup \{l_{i-1}\}$} \\
      \rho & \text{else},
      \end{cases}
\end{align}}
\textcolor{blue}{where $l_{i+1} \in C_\omega(l_i)$}.
The sampling strategy is similar to a second order Markov chain \cite{mark}, in which the probability of adding the next node is not only influenced by the current node but also the previous node. \textcolor{blue}{A low} value of \textcolor{blue}{$\rho$} boosts the rate of sampling an element from $C_\omega(l_{i-1})$. On the contrary, a high value of $\rho$ controls the probability of exploring a node far from $l_{i-1}$. \textcolor{blue}{Higher $\rho$} allows sampling a node in $C_\omega(l_i)$ but not in $C_\omega(l_{i-1})$. If \textcolor{blue}{$\rho=1$}, Eq. \ref{eq:s9} is equivalent to Eq. \ref{eq:s8}.

In the $\omega$th condition, we generate $K$ random node sequences starting from each node. Repeating the process for all the $M$ conditions, we \textcolor{blue}{obtain} a corpus $T$ containing $KNM-KZ$ node sequences, where $Z$ represents the number of missing values.

\textcolor{blue}{The goal of generating random node sequences is to feed the corpus $T$ to a three-layer neural networks to obtain node vectors} \cite{lan1,lan2,lan3,barkan,bnb}. \textcolor{blue}{Please refer to Appendix \ref{apA} for more information about the neural network model.}

\subsection{\textcolor{blue}{Construct network from trained node vectors}} \label{IIC} 
After training the neural network model, \textcolor{blue}{we obtain $N$ vectors} for the $N$ nodes. With the node vectors, we can predict \textcolor{blue}{relationships between the nodes, visualize the global structures of the nodes, and construct a corresponding network.}

A conventional way to select the edges is by global thresholding the cosine similarities to filter out weak links and obtain a backbone of the underlying network. Globally thresholding edges (GTE) is widely used in determining gene co-expression networks \cite{horvath3}. However, the drawback of the GTE is that some nodes could be isolated from the network if the threshold is high. \textcolor{blue}{Though we can force isolated nodes connected to some other nodes, the degree distribution of the constructed network is still affected by the selection of threshold, meaning that the roles of nodes in the network are sensible to the choice of threshold.} To avoid these issues, we propose a R\'enyi entropy-based method (REM) to \textcolor{blue}{extract a network from the trained node vectors} \cite{lee, hill, entropy3, entropy4}. 

Once we have the node vectors, we can compute the cosine similarity to quantify the connection strength for each pair of nodes. Here, we \textcolor{blue}{define $S^0(v_i)$ as the initial neighbor set of node $v_i$.} \textcolor{blue}{$S^0(v_i)$ is composed of nodes} that are positively similar to $v_i$, i.e., $S^0(v_i)=\{v_j:s(v_i, v_j)>0\}$, where $s(v_i, v_j)$ is the cosine similarity between $v_i$ and $v_j$. \textcolor{blue}{The network constructed from $S^0(v_i),\ \forall i < N$ is not helpful in real applications because most node pairs are weakly connected.}

Inspired by the application of entropy in ecology, we regard the \textcolor{blue}{nodes in the set $S^0(v_i)$ are the states of the system $v_i$}. Then, we associate each state with a probability, which is computed from the similarity values, such that:
\begin{align}
\label{eq:s14}
\overline{s}_i^0(v_j)=\frac{s(v_i, v_j)}{\sum_{v_j\in S^0(v_i)}{s(v_i, v_j)}}.
\end{align}

In information theory, entropy depicts the diversity and randomness of a system \cite{entropy2}. The R\'enyi entropy for node $v_i$ with order $\alpha$ is 
\begin{align}
\label{eq:s15}
H_{\alpha}^1(v_i)=\frac{1}{1-\alpha} \ln \sum_{v_j \in S^0(v_i)} (\overline{s}_i^0(j))^{\alpha},
\end{align}
where $\alpha>0$. Note that the R\'enyi entropy converges to the Shannon entropy in the case $\alpha\to1$, i.e., $H_\alpha^1(v_i)=\sum_{v_j \in S^0(v_i)} \overline{s}_i^0(v_j) \log \overline{s}_i^0(v_j)$. For any $\alpha$, the entropy $H_\alpha^1(v_i)$ varies from zero to $\ln |S^0(v_i)|$. In the case of a certain event, i.e., \textcolor{blue}{ $\exists\ v_j \in S^0(v_i)$, where $\overline{s}_i^0(v_j)=1$ and $H_\alpha(v_i)=0$}. Conversely, the entropy $H_\alpha^1(v_i)=\ln |S^0(v_i)|$ when $\overline{s}_i^0(v_j)$ follows a uniform distribution. The diversity index $D_{\alpha}^1(v_i)$ is
\begin{align}
\label{eq:s16}
D_{\alpha}^1(v_i)&=\exp(H_\alpha^1(v_i)) = \left(\sum_{v_j \in S^0(v_i)} (\overline{s}_i^0(v_j))^{\alpha}\right)^{1/(1-\alpha)},
\end{align}
which is also known as the Hill numbers \cite{hill}. It is unsurprising that $\overline{s}_i^0(v_j)$ is not \textcolor{blue}{uniformly distributed}, and $H_\alpha^1(v_i) \in [0,\ \ln |S^0(v_i)|]$. In ecology, the diversity index quantifies the abundance of species in a community. The diversity index approaches the total number of species when the species are equally abundant \textcolor{blue}{and approaches} one if there is a dominant species. In Eq. \ref{eq:s16}, the order $\alpha$ influences the sensitivity of the diversity index. Increasing $\alpha$ strengthens the weights of the most abundant species. \textcolor{blue}{That is, higher $\alpha$ allows us to select the more abundant species, while lower $\alpha$ will detect more species.} Therefore, we can use $\alpha$ to control the number of neighbors of node $v_i$.

We pick $\nint{D_{\alpha}^1(v_i)}$ nodes that have highest similarities from $S^0(v_i)$ as the effective number of neighbors of node $v_i$. Then, the selected neighbors compose a new neighbor set $S^1(v_i)$. \textcolor{blue}{The nodes in $S^1(v_i)$ are more strongly connected to $v_i$ than the nodes in $S^0(v_i)$. The network constructed from $S^1(v_i), \forall i \leq N$ is denser than that constructed from $S^0(v_i), \forall i \leq N$. We can repeatedly run Eqs. \ref{eq:s14}, \ref{eq:s15}, and \ref{eq:s16} to obtain a network with desired edge density. Assume} $\nint{D_{\alpha}^k(v_i)}$ is the diversity index of $k$th iteration. Then, we have
\begin{align}
\label{eq:s17}
S^k(v_i)&=\{v_j\in S^{k-1}(v_i):\nonumber\\
&|\{v_z\in S^{k-1}(v_i):s(v_i,v_j)<s(v_i,v_z)\}|<\nint{D_{\alpha}^k(v_i)}\},
\end{align}
where $k\geq 1$. In each iteration, $\nint{D_{\alpha}^k(v_i)}$ nodes with highest similarity values are selected as the neighbors of $v_i$. Intuitively, the REM can filter out weak links for $v_i$, \textcolor{blue}{and the remaining nodes $S^k(v_i)$ are the most \emph{meaningful} neighbors of $v_i$.}

\textcolor{blue}{In real networks, leaf nodes are those connected to a small number of others, while hubs have many neighbors. Considering the property of entropy \cite{entropy3, entropy4}, the size of the resulting neighbor set $S^k(v_i)$ is relatively small if the similarity value distribution of $S^0(v_i)$ is right-skewed. On the contrary, the size of $S^k(v_i)$ is much larger if the similarity value distribution of $S^0(v_i)$ is left-skewed \cite{lee, hillskew1, hillskew1}. That is, the role of node $v_i$ in the resulting network is related to the similarity value distribution.}

\section{Results}
The method we have presented falls in the category of unsupervised learning. In this section, \textcolor{blue}{we use both synthetic and real data to evaluate the performance of the proposed approach in recovering global and local structures in terms of feature learning  and network reconstruction.}

\subsection{\textcolor{blue}{Feature learning}}

\paragraph{\textcolor{blue}{Synthetic data.}} \textcolor{blue}{In this part, we used two case studies with $N_1=5000$ and $N_2=5500$ nodes to evaluate the performance of the proposed approach in recovering a global structure.} \textcolor{blue}{The nodes in the two case studies} are measured in six conditions ($M_1$, $M_2$, $\cdots$, $M_6$) and distributed in five communities ($G_1$, $G_2$, $\cdots$, $G_5$). \textcolor{blue}{The first case study has five communities of equal size, i.e., each group has 1,000 nodes. The five communities in the second case study have respectively $G_1=1000$, $G_2=1500$, $G_3=500$, $G_4=750$, and $G_5=1750$ nodes (The sizes of the communities are chosen randomly).} \textcolor{blue}{Note that the sixth condition is a perturbation.} In each condition, nodes in the same community are assigned random values from one of the intervals: $A=[1, 100]$, $B=[101, 200]$, $C=[201, 300]$, $D=[301, 400]$, $E=[401, 500]$ , and $R=[1, 500]$. \textcolor{blue}{In this work, we created four datasets for each case study per the tables in Appendix \ref{apB}. In Table. \ref{tab:tabd1} ($Data.1$), $G_1$ and $G_2$ are adjacent but not overlapped. In Tables \ref{tab:tabd2} ($Data.2$), \ref{tab:tabd3} ($Data.3$), and \ref{tab:tabd4} ($Data.4$), nodes from $G_1$ and $G_2$ are respectively assigned with values from two, three, and four same intervals, as shown in bold fonts.} The relative distance between $G_1$ and $G_2$ is expected to decrease with respect to the increase of the number of overlapped intervals.

\paragraph{\textcolor{blue}{Experimental results.}} \textcolor{blue}{Based on the approach introduced in Section II, we generated context sets with a tolerance of} $\delta_i^\omega = 0.1v_i(\omega)$ ($\beta_\omega=0.1$). \textcolor{blue}{In the experiments,} we generated $K=10$ random node sequences of length $l=80$ starting from each node in each of the six conditions. Consequently, the corpus $T_1$ and $T_2$ consist of $300000$ and $330000$ node sequences, respectively. In the neural network, we set the node vector dimension to $d=128$.

\begin{figure}[ht]
\centering
\includegraphics[width=0.49\linewidth]{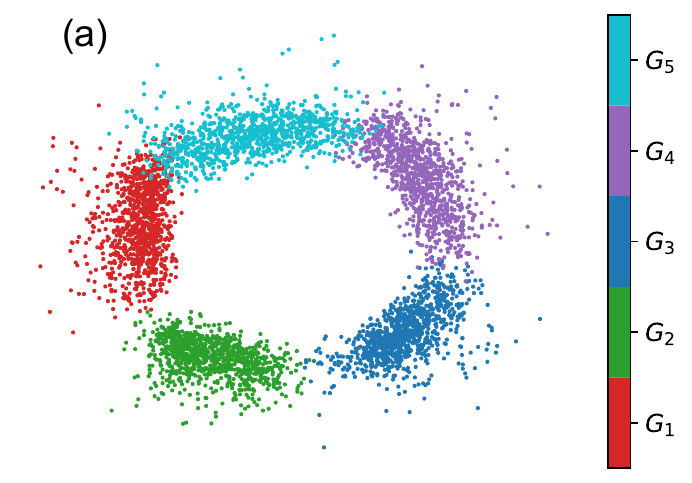}
\includegraphics[width=0.49\linewidth]{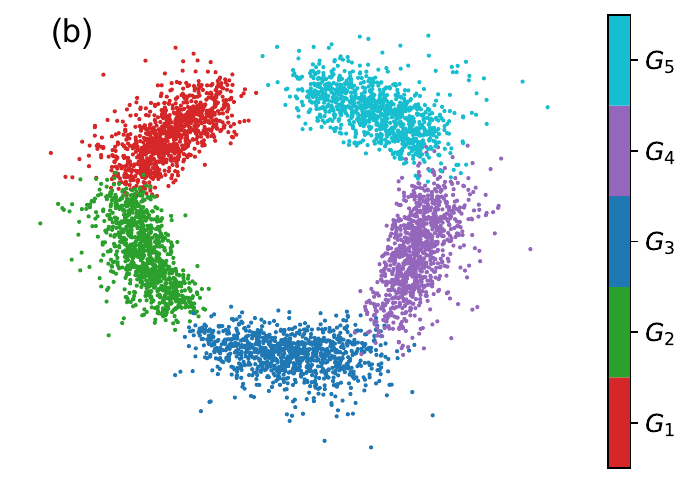}
\includegraphics[width=0.49\linewidth]{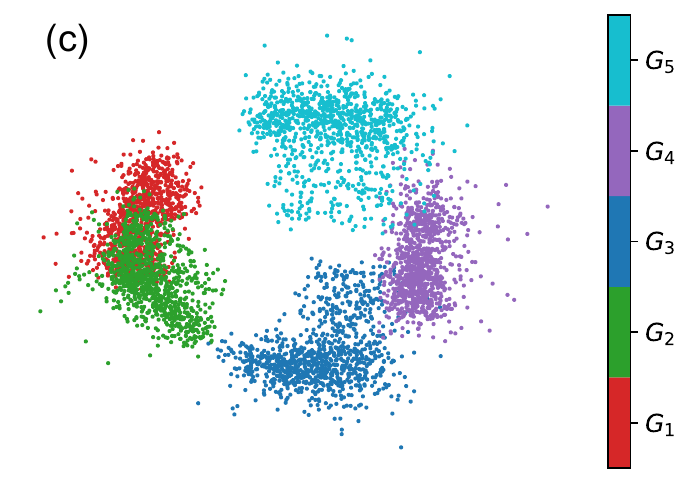}
\includegraphics[width=0.49\linewidth]{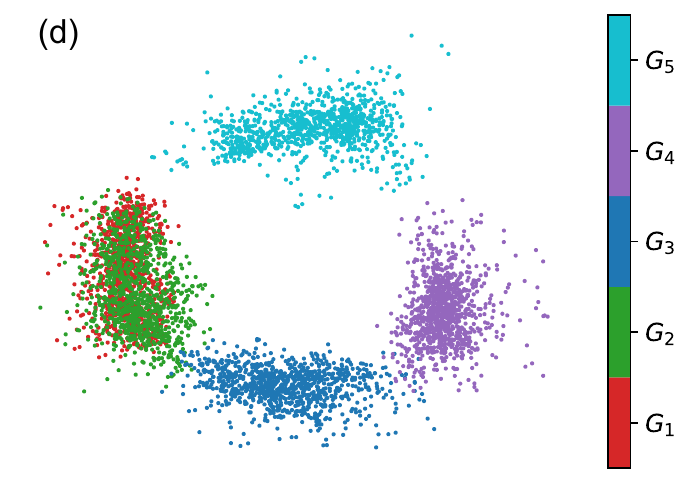}
\caption{\label{fig:fig1} The node vectors trained from the \textcolor{blue}{first case study} are visualized via PCA. The five communities have an equal number of nodes. \textcolor{blue}{Panels (a) to (d) are respectively the training results of $Data.1$ to $Data.4$.}}
\end{figure}

\begin{figure}[ht]
\centering
\includegraphics[width=0.49\linewidth]{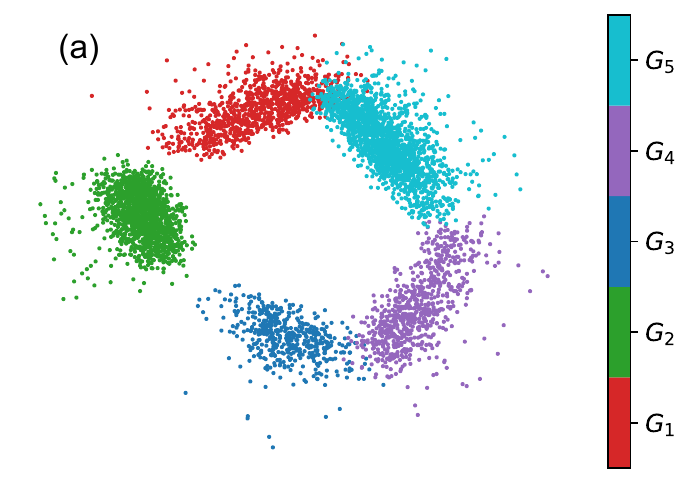}
\includegraphics[width=0.49\linewidth]{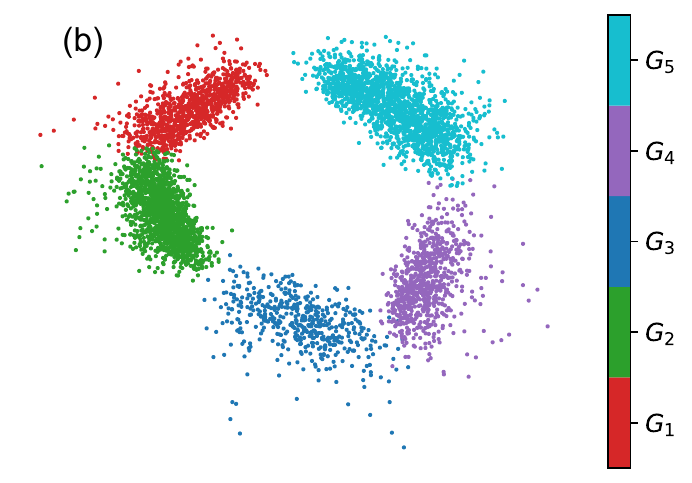}
\includegraphics[width=0.49\linewidth]{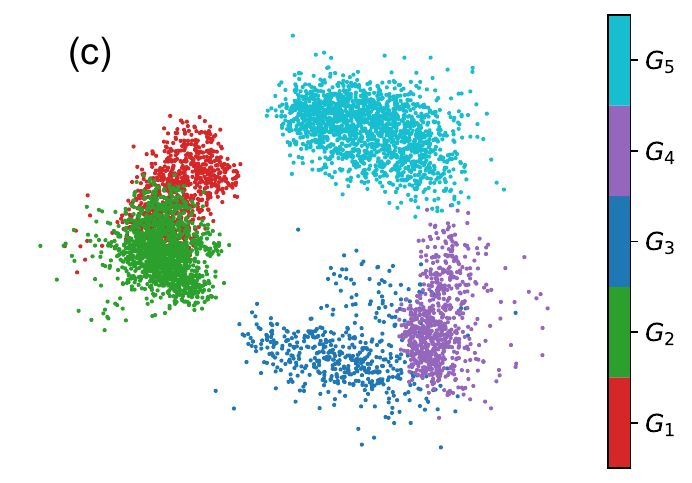}
\includegraphics[width=0.49\linewidth]{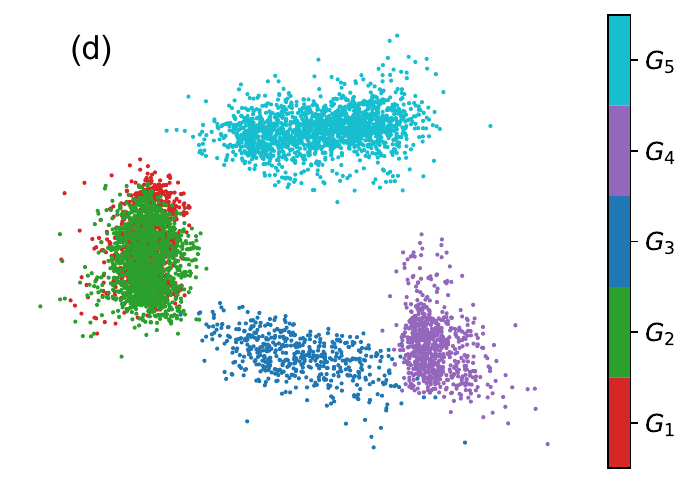}
\caption{\label{fig:fig2} The node vectors trained from the second \textcolor{blue}{case study} are visualized via the PCA. The five communities have 1000, 1500, 500, 750, and 1250 nodes. \textcolor{blue}{Panels (a) to (d) are respectively the training results of $Data.1$ to $Data.4$.}}
\end{figure}

To evaluate the training results qualitatively, we mapped the trained node vectors to a 2D plane via the Principle component analysis (PCA) \cite{pca, filz}. In Fig. \ref{fig:fig1}(a) and \ref{fig:fig2}(a), the nodes from the same communities are mapped to the same areas, \textcolor{blue}{meaning that the proposed method can recover the global structure of the dataset.} \textcolor{blue}{Note that nodes in $G_1$ and $G_2$ are assigned to values from two, three, and four overlapped sub-intervals from $Data.2$ to $Data.4$ (see the details in Appendix \ref{apB}). That is, the distance between $G_1$ and $G_2$ is assumed to be decreasing for $Data.2$, $Data.3$, and $Data.4$.} In Fig. \ref{fig:fig1}(b) and \ref{fig:fig2}(b), we observed the relative distance between $G_1$ and $G_2$ was closer than that in Fig. \ref{fig:fig1}(a) and \ref{fig:fig2}(a). Similarly, the relative distance between $G_1$ and $G_2$ was even closer in Fig. \ref{fig:fig1}(c) and \ref{fig:fig2}(c), and the two communities were almost merged in Fig. \ref{fig:fig1}(d) and \ref{fig:fig2}(d). The results for the two case studies (eight datasets in total) demonstrated that the node vectors can reflect the relative distances of the node communities, which are affected by the number of overlapped sub-intervals.

\begin{table}[h]
\caption{\label{tab:tab1}The relative distance between $G_1$ and $G_2$ \textcolor{blue}{and classification accuracy}}
\begin{ruledtabular}
\begin{tabular}{ccccc}
&\multicolumn{2}{c}{First case study}&\multicolumn{2}{c}{Second case study}\\
\textrm{Data.}&
\textrm{Distance}&
\textrm{Accuracy(\%)}&
\textrm{Distance}&
\textrm{Accuracy(\%)}\\
\colrule
1&2.18&99.98&2.14&100\\
2&1.82&99.96&1.86&99.98\\
3&1.52&99.96&1.49&99.90\\
4&1.18&79.84&1.16&77.92\\
\end{tabular}
\end{ruledtabular}
\end{table}

In order to quantitatively show the results, we computed the distance between $G_1$ and $G_2$. To this end, \textcolor{blue}{we calculated the cosine distance (1-cosine similarity) between node pairs.} The distance between $G_1$ and $G_2$ was computed as the summation of all possible node pairs between the two communities. For example, the cosine distance between $G_1$ and $G_2$ is
\begin{align}
\label{eq:sr31}
Dis (G_1,G_2 )=\sum_{v_i\in G_1, v_j\in G_2} 1-s(v_i, v_j).
\end{align}
\textcolor{blue}{Then, we calculate the relative distance between $G_1$ and $G_2$ as:}
\begin{align}
\label{eq:sr32}
RelaDis (G_1, G_2)=\frac{Dis(G_1,G_2 )*Dis(G_1,G_2 )}{Dis(G_1, G_1)*Dis(G_2,G_2)}.
\end{align}

\textcolor{blue}{Additionally, we perform the simple \textit{K}-means clustering method \cite{kmean} to classify the trained node vectors into five communities. The classification results are compared to the ground-truth communities.}

\textcolor{blue}{The relative distances between $G_1$ and $G_2$ and classification results are shown in Table. \ref{tab:tab1}. We observe that the cosine distance between $G_1$ and $G_2$ is decreasing for $Data.1$ to $Data.4$, in accord with the visualizations in Fig. \ref{fig:fig1} and Fig. \ref{fig:fig2}. Specifically, the distance is close to one for $Data.4$, which suggests that the two communities almost merged. The classification results also agree with the visualization. The classification accuracy is above 99\% for $Data.1$, $Data.2$, and $Data.3$, and the classification accuracy has dropped significantly in $Data.4$ since the two communities are almost overlapped, and the nodes from the two communities are falsely classified. From a global view, the node vectors can recover the mesoscopic structure of the nodes. In the following experiments, we will only use the first case study to conduct further analysis.}

\begin{table}[h]
\caption{\label{tab:tab2}\textcolor{blue}{The influence of missing values on the distance between $G_1$ and $G_2$ and classification accuracy}}
\begin{ruledtabular}
\begin{tabular}{ccccc}
&\multicolumn{2}{c}{10\% missing values}&\multicolumn{2}{c}{20\% missing values}\\
\textrm{Data.}&
\textrm{Distance}&
\textrm{Accuracy(\%)}&
\textrm{Distance}&
\textrm{Accuracy(\%)}\\
\colrule
1&2.19&100&2.16&99.92\\
2&1.77&99.92&1.76&99.72\\
3&1.49&99.64&1.47&98.76\\
4&1.15&79.68&1.14&79.96\\
\end{tabular}
\end{ruledtabular}
\end{table}

\begin{figure}[ht]
\centering
\includegraphics[width=0.49\linewidth]{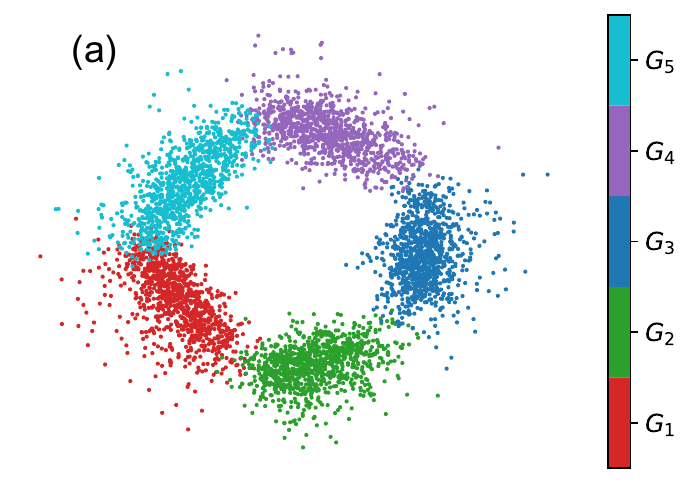}
\includegraphics[width=0.49\linewidth]{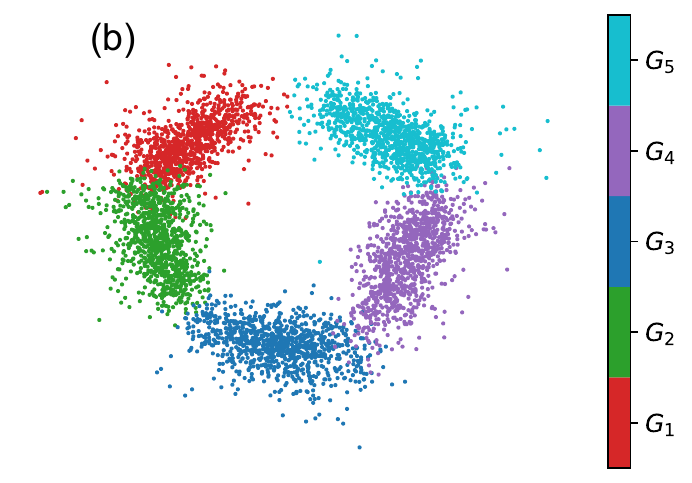}
\includegraphics[width=0.49\linewidth]{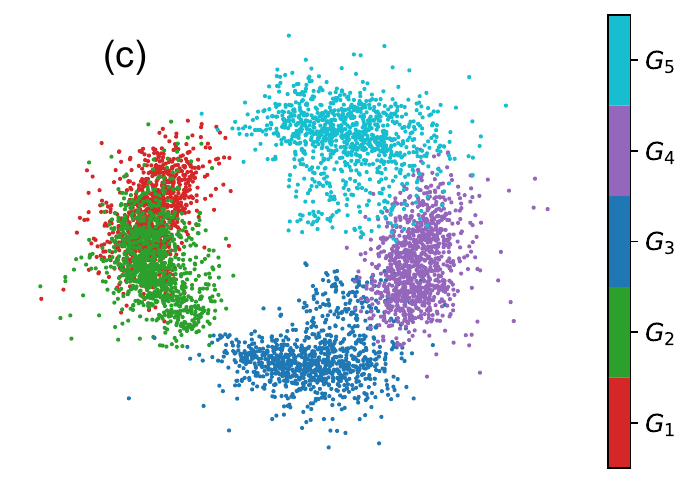}
\includegraphics[width=0.49\linewidth]{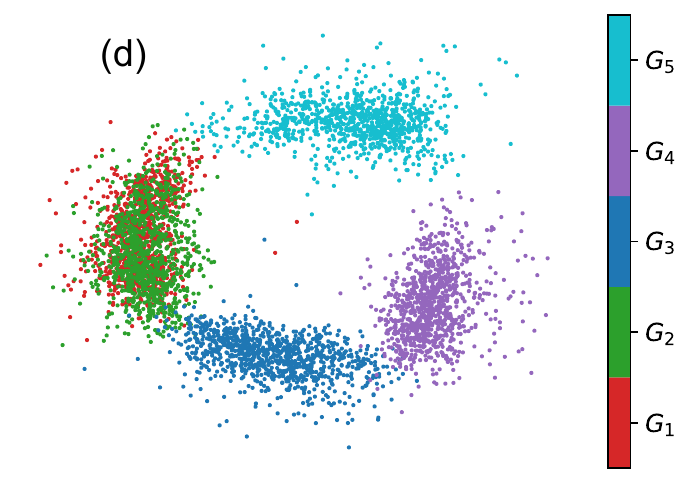}
\caption{\label{fig:fig3} Visualization of node vectors trained from the first example with 20\% missing values. Panels (a) to (d) are respectively the training results of $Data.1$ to $Data.4$.}
\end{figure}

\textcolor{blue}{To study the influence of missing values,} we generate incomplete datasets by randomly removing 10\% and 20\% of \textcolor{blue}{values from each condition.} \textcolor{blue}{We use the same parameters to train the neural network, and the results are shown in Table \ref{tab:tab2}. It can be observed that} the relative distances between $G_1$ and $G_2$ \textcolor{blue}{and the classification accuracies are not significantly affected by the missing values.} In Fig. \ref{fig:fig3}, the visualization shows that \textcolor{blue}{the global structure of the nodes} can still be recovered even when 20\% data have been removed randomly. \textcolor{blue}{Therefore, the results suggest the proposed method is robust to missing values.}

\begin{table}[h]
\caption{\label{tab:tab3}\textcolor{blue}{The relative distance between $G_1$ and $G_2$ and classification accuracy w.r.t. the variation of $\beta_\omega$}}
\begin{ruledtabular}
\begin{tabular}{ccccccccc}
&\multicolumn{4}{c}{Distance}&\multicolumn{4}{c}{Accuracy(\%)}\\
\textrm{Data.}&
\textrm{0.05}&
\textrm{0.1}&
\textrm{0.15}&
\textrm{0.2}&
\textrm{0.05}&
\textrm{0.1}&
\textrm{0.15}&
\textrm{0.2}\\
\colrule
1&2.08&2.18&2.22&2.16&100&99.98&99.86&99.86\\
2&1.78&1.82&1.82&1.76&99.98&99.96&99.86&99.88\\
3&1.49&1.52&1.48&1.47&99.92&99.96&99.06&99.84\\
4&1.15&1.18&1.16&1.14&80.72&79.84&80.86&81.34\\
\end{tabular}
\end{ruledtabular}
\end{table}

\textcolor{blue}{The training results are robust to the choice of training parameters. In the generation of node sequences, we assigned different values to $\beta_\omega$ to control the size of the context set $\delta_\omega^i$. The node sequences are trained using the neural network model with $\rho=1$. Similarly, we computed the relative distances between $G_1$ and $G_2$ and the classification accuracy. Table \ref{tab:tab3} shows that the relative distances are at the same levels for the same datasets, and the classification accuracies are not significantly affected by $\beta_\omega$, meaning that the global structure is still maintained. Thus, the choice of $\beta_\omega$ has limited influence on the embedded node vectors.}

\textcolor{blue}{To compare the proposed approach with the widely used correlation approach \cite{horvath3, reverter}, we generated four networks from the synthetic data and trained the networks with semi-supervised learning algorithms to obtain node vectors. First, the values of each condition are normalized with the z-score:
\begin{align}
\label{eq:c1}
\overline{v}_i(\omega)=\frac{v_i (\omega)-\mu_\omega}{\sigma_\omega}
\end{align}
where $\mu_\omega$ is the mean of all the values in the  $\omega$th condition, $\sigma_\omega$ is the standard deviation, and $\overline{v}_i(\omega)$ is the normalized expression value.}

\textcolor{blue}{The Pearson correlation coefficient (PCC) of any two nodes is:
\begin{align}
\label{eq:c2}
r_{(v_x,v_y)}=\frac{\sum_{i=1}^M (\overline{v}_x(\omega)-\overline{v}_x)(\overline{v}_y(\omega)-\overline{v}_y)}{\sqrt{\sum_{i=1}^M (\overline{v}_x(\omega)-\overline{v}_x)^2}\sqrt{\sum_{i=1}^M (\overline{v}_y(\omega)-\overline{v}_y)^2}}
\end{align}
where $r_{(v_x, v_y)}$ is the PCC between node $v_x$ and $v_y$, and $\overline{v}_x$ is the mean of node $v_x$ across the $M$ conditions. The PCC measures how much the two genes are related \cite{horvath3, cancer}. In this experiment, we did not consider the missing value problem, which could substantially influence the correlation coefficients according to the results in \cite{reverter}. The edges are selected by thresholding correlation coefficients, such that PCC $\geq 0.95$ \cite{cancer, bon1}. All four networks have edge densities above 5\%, as shown in Table \ref{tab:tab4}}.

\begin{table}[h]
\caption{\label{tab:tab4}\textcolor{blue}{The relative distance between $G_1$ and $G_2$ and the classification accuracy of the PCC networks}}
\begin{ruledtabular}
\begin{tabular}{ccccc}
\textrm{Data.}&\textrm{Edge density}&\textrm{Distance}&\textrm{Accuracy(\%)}\\
\colrule
1&6.30\%&6.13&100\\
2&5.81\%&3.64&81.30\\
3&5.21\%&2.66&80.82\\
4&5.73\%&1.17&64.26\\
\end{tabular}
\end{ruledtabular}
\end{table}

\begin{figure}[ht]
\centering
\includegraphics[width=0.495\linewidth]{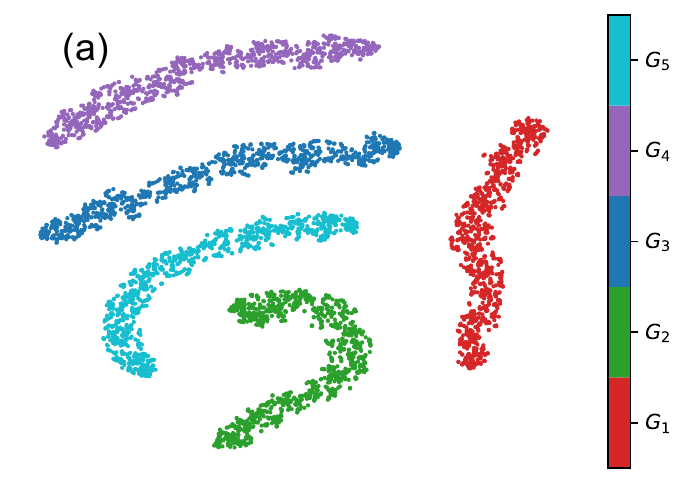}
\includegraphics[width=0.475\linewidth]{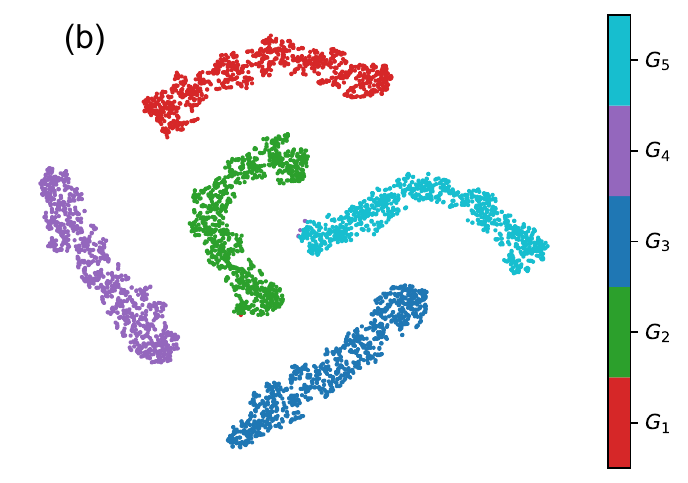}
\includegraphics[width=0.492\linewidth]{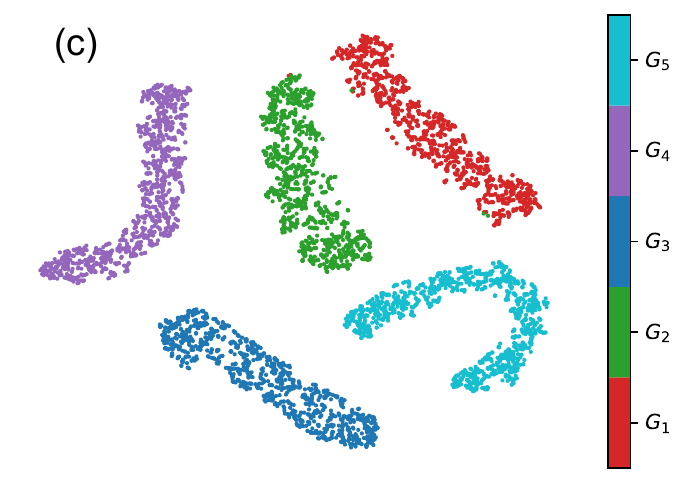}
\includegraphics[width=0.492\linewidth]{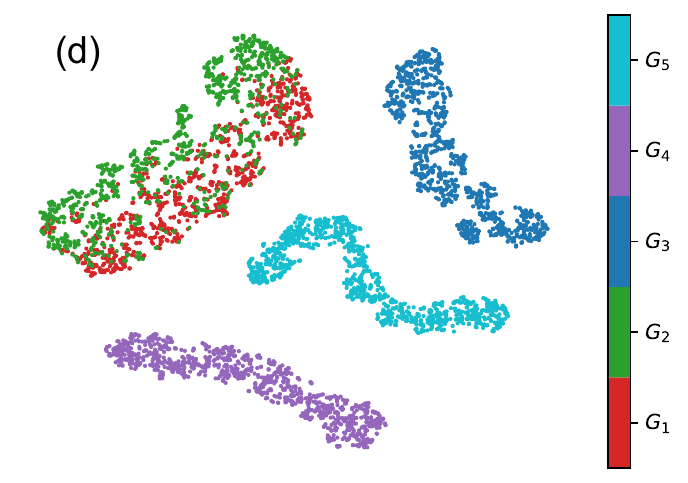}
\caption{\label{fig:fig4} \textcolor{blue}{The visualization of node vectors trained from the Pearson correlation network. Panels (a) to (d) are respectively $Data.1$ to $Data.4$.}}
\end{figure}

\textcolor{blue}{To study the properties of nodes, different methodologies are used to determine node vectors from network structure \cite{vec1, vec2, vec3, ball}. Here, we used the node2vec method introduced in \cite{vec1} to obtain node vectors from the constructed networks since the approach has shown outstanding performance in reconstructing networks. Similarly, we computed the relative distances between $G_1$ and $G_2$ from the trained node vectors, and the results are shown in Table \ref{tab:tab4}. We observed that the relative distances between the two communities for the first three networks are much higher than in our method (Table \ref{tab:tab1}). In the synthetic datasets, $G_1$ and $G_2$ are assumed to partially overlap. However, this characteristic is not recovered from trained node vectors per the visualization of Fig. \ref{fig:fig4}. In the PCC method, errors could be introduced in data normalization, network construction, and feature learning, which consequently influence the accuracy of trained node vectors. As a comparison, our proposed approach trains the node vectors directly from the raw data.}

\textcolor{blue}{More experimental results on the choice of $\rho$ can be found in Appendix \ref{apC}.}

\begin{figure}[ht]
\centering
\includegraphics[width=0.49\linewidth]{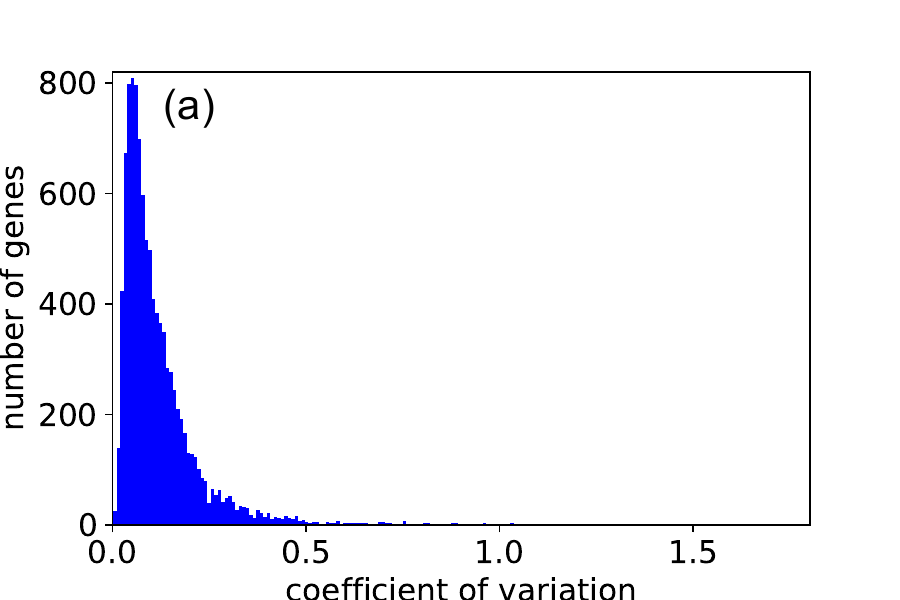}
\includegraphics[width=0.49\linewidth]{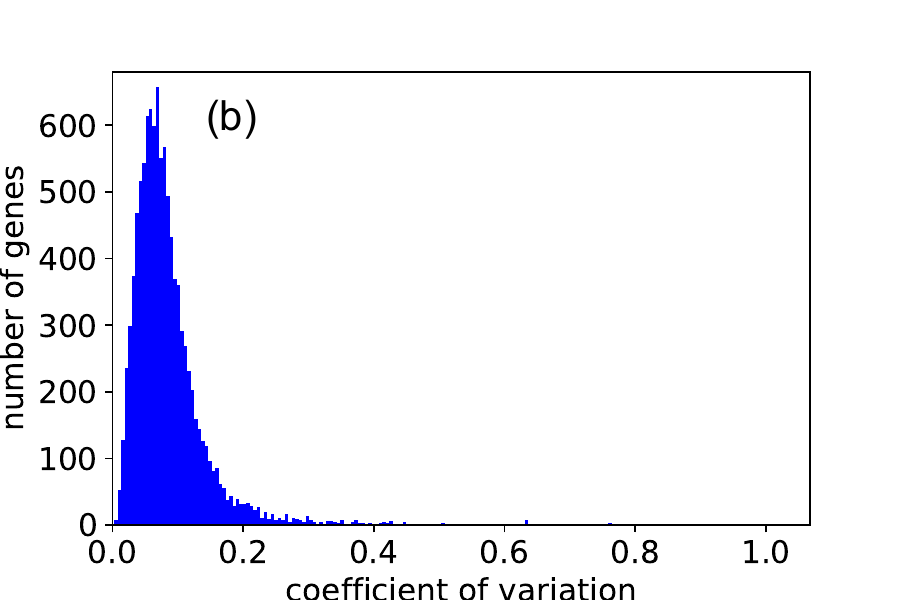}
\includegraphics[width=0.49\linewidth]{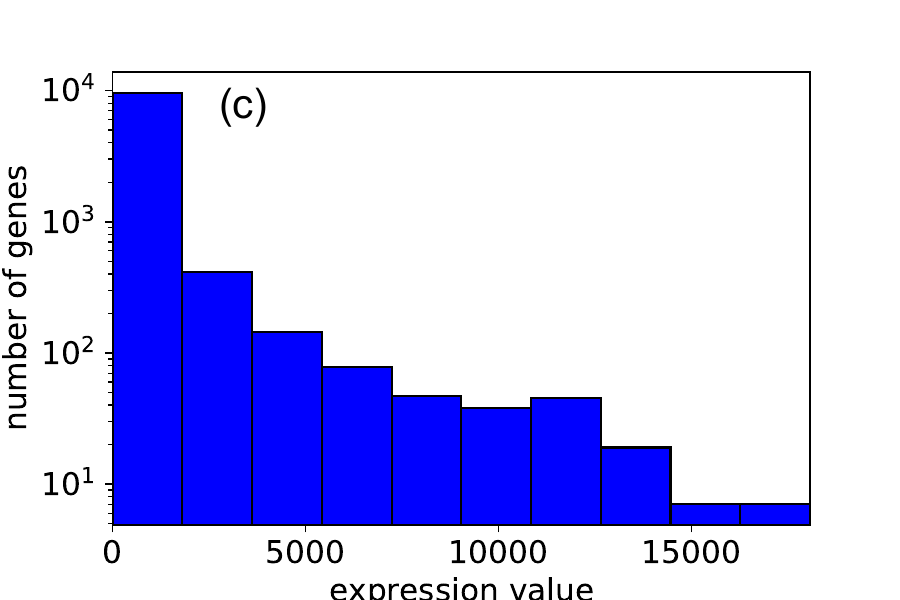}
\includegraphics[width=0.49\linewidth]{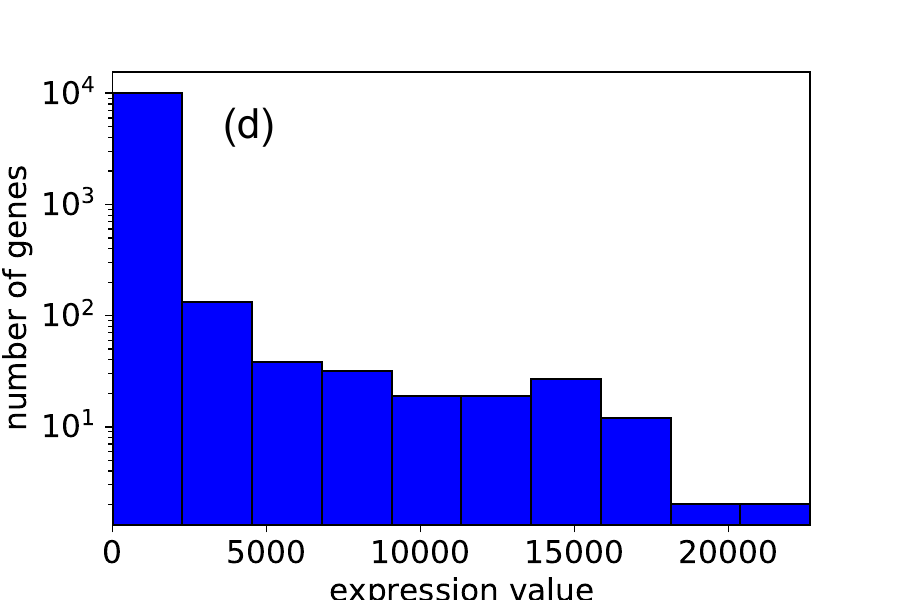}
\caption{\label{fig:fig5} The properties of the two real datasets. Panels (a) (first dataset) and (b) (second dataset) are the distributions of the $CV$ of the expression values. Panels (c) (first dataset) and (d) (second dataset) are the distributions of means of the expression values.}
\end{figure}

\paragraph{\textcolor{blue}{Real data.}} We used two real \emph{Anopheles gambiae} gene expression datasets \cite{bob, kuangano} to show that the learned node vectors can capture the local structure of the nodes. The first dataset consists of 10,433 \emph{Anopheles gambiae} genes measured in time series after desiccation stress (five conditions) \cite{wang}. The five measurements (conditions) of each gene are almost at the same level, and the distributions of the \emph{coefficient of variation} ($CV$) and means of the 10,433 genes are shown in Fig. \ref{fig:fig5}(a) and (c). The second dataset measures the gene expression values after mating \cite{rogers}, consisting of four measurements (also in time series). The distributions of the $CV$ and means are shown in Fig. \ref{fig:fig5}(b) and (d).     

\textcolor{blue}{In Fig. \ref{fig:fig5}(a) and (b), we observe that the $CV$s of most genes are at a low level.} Therefore, we can set the tolerance $\delta_i^\omega$ as the average $CV$ of all the nodes, such that
\begin{align}
\label{eq:s18}
\overline{CV}= \frac{1}{N}\sum_i \frac{\sigma_i}{m_i},
\end{align}
where $m_i$ is the mean value of gene $v_i$, $\sigma_i$ is the standard deviation, and $\frac{\sigma_i}{m_i}$ is the $CV$ of gene $v_i$. The $\overline{CV}$s of the two data sets are respectively 0.086 and 0.12. Therefore, we set $\beta_\omega=\overline{CV}$.

\begin{figure}[ht]
\centering
\includegraphics[width=0.49\linewidth]{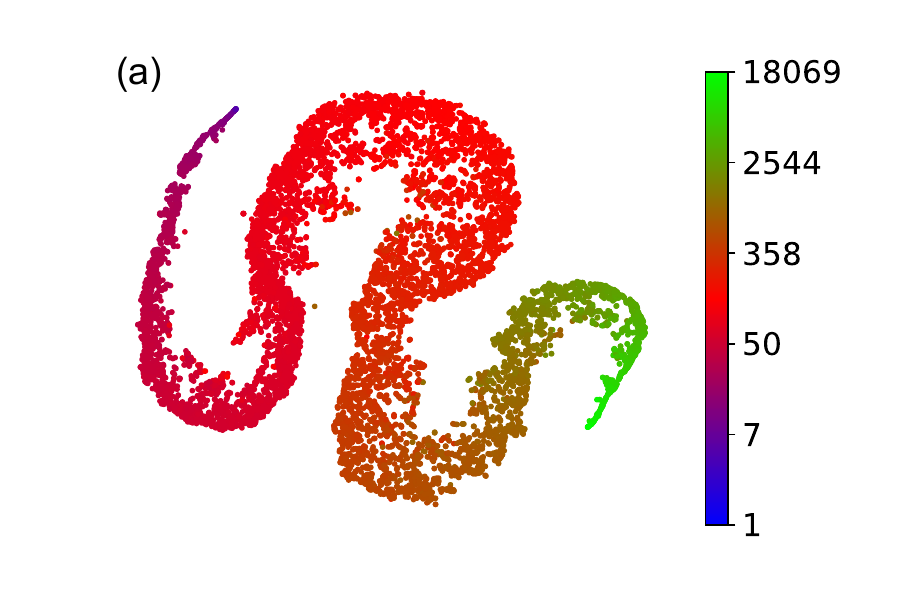}
\includegraphics[width=0.49\linewidth]{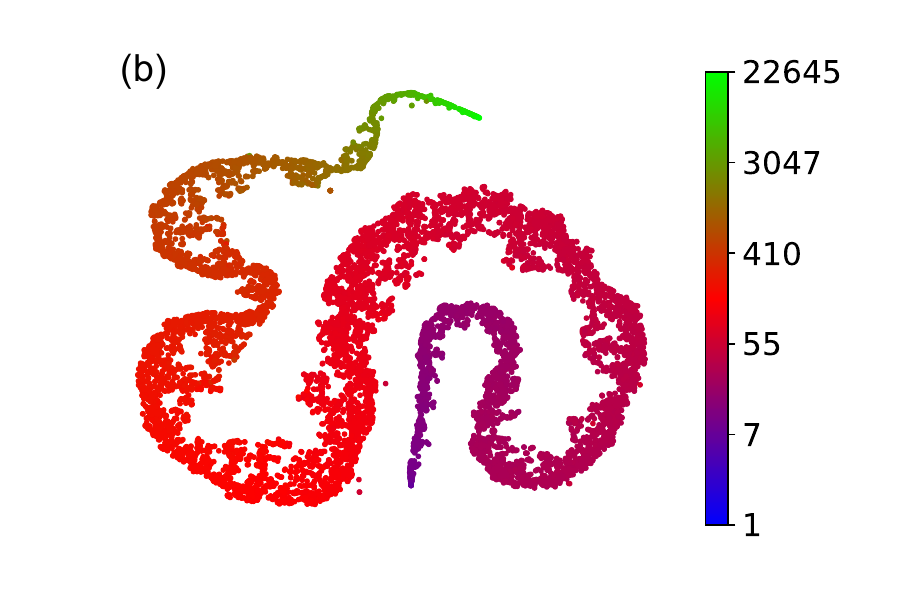}
\includegraphics[width=0.49\linewidth]{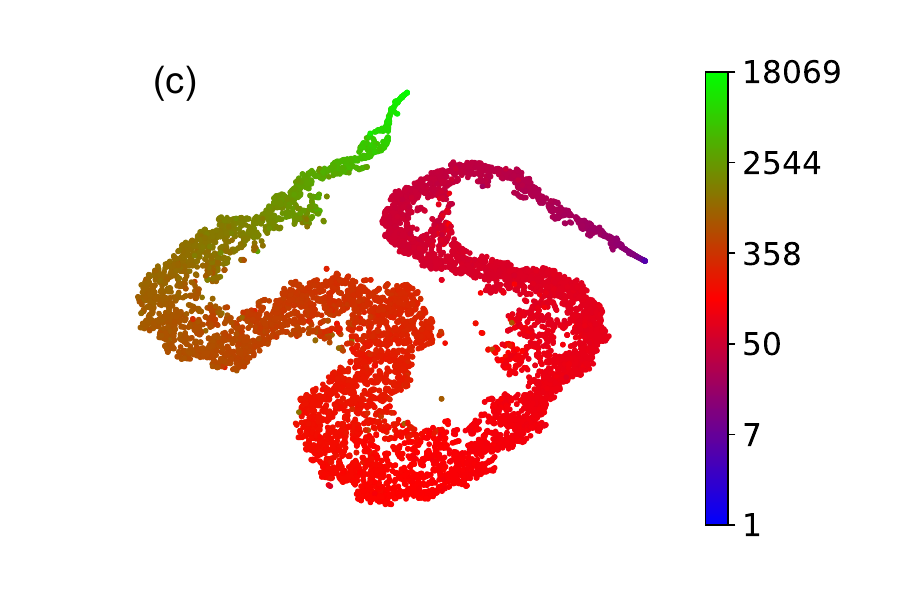}
\includegraphics[width=0.49\linewidth]{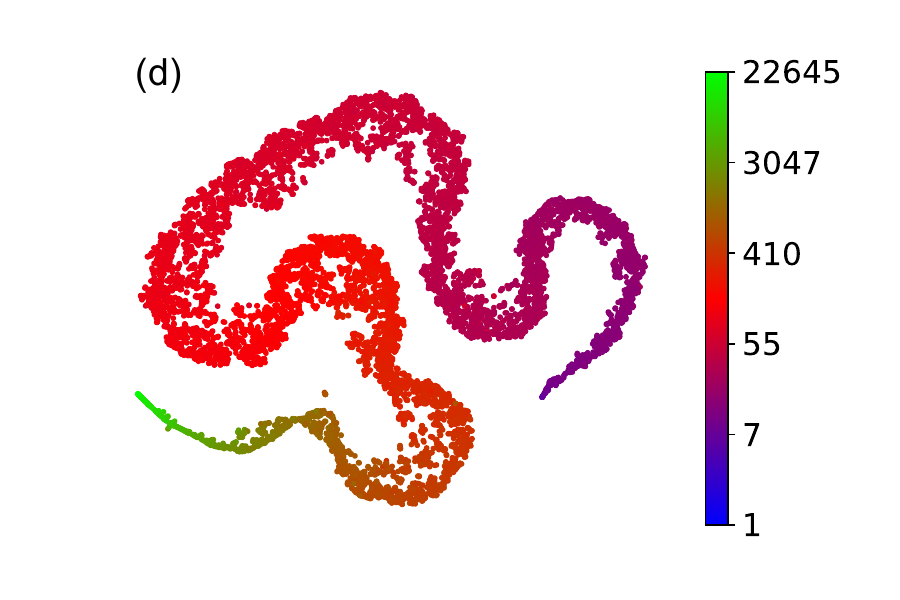}
\caption{\label{fig:fig6} Visualization of the gene vectors via t-SNE, the genes are colored by the expression values. \textcolor{blue}{Panels (a) (first dataset) and (b) are the visualizations of node vectors trained from raw data. Panels (c) (first dataset) and (d) (second dataset) are the visualizations of node vectors trained from the incomplete datasets with 20\% values randomly removed from each condition.}}
\end{figure}

\paragraph{\textcolor{blue}{Experimental results.}} \textcolor{blue}{The trained node vectors are visualized via the t-distributed stochastic neighbor embedding (t-SNE) method \cite{maaten} in Fig. \ref{fig:fig6}.} \textcolor{blue}{The t-SNE constructs probability distribution over pairs of vectors and does not retain the distances of node pairs, but their probabilities. Therefore, the t-SNE approach has better performance in preserving local structure. In Fig. \ref{fig:fig6}, we can observe that the genes with similar expression values are mapped closer, even when 20\% of values have been removed from each condition.}

 \begin{figure}[ht]
\centering
\includegraphics[width=0.98\linewidth]{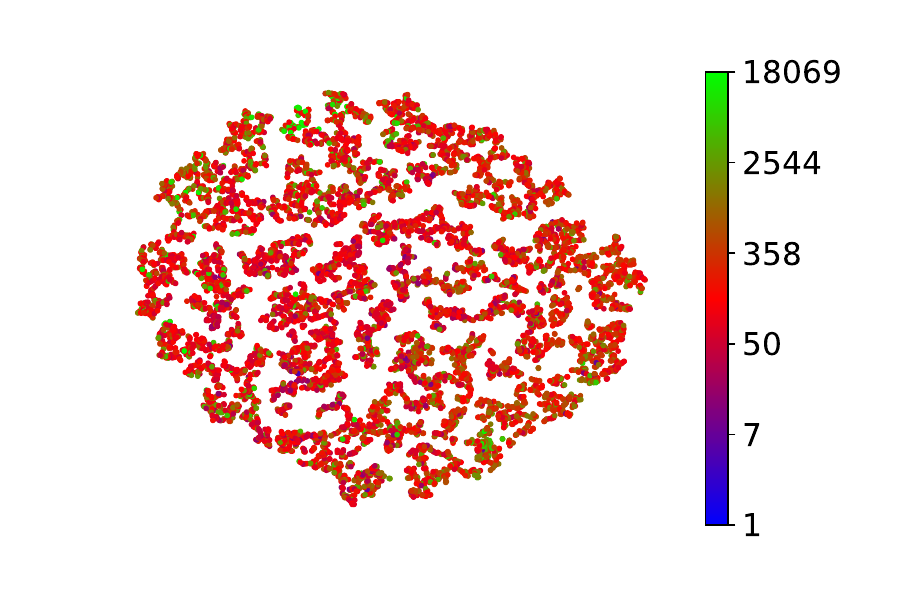}
\caption{\label{fig:fig7} \textcolor{blue}{The visualization of node vectors trained from the PCC network.}}
\end{figure}

\textcolor{blue}{As a comparison, we construct a PCC network for the first real data. The raw expression values are rescaled with log2 \cite{bob, cancer} and normalized per Eq. \ref{eq:c1} (the distribution of the raw data is heterogeneous as shown in Fig. \ref{fig:fig5}(c)). Then, a PCC network is constructed by thresholding the edges with a threshold PCC$\geq 0.95$ \textcolor{blue}{(the network is not sparse)}. The resulting network consists of 756,330 edges. Similarly, the node vectors are obtained by training the node2vec model. In Fig. \ref{fig:fig7}, we observed that nodes are distributed randomly in the 2D plane, suggesting that nodes with close values are not mapped to the same area. For example, the expression values of the two genes AGAP004677 and AGAP012093 are respectively [2764, 2869, 3276, 3690, 3671] and [129, 149, 184, 221, 265], and it is apparent that the expression values of the two genes are at different levels. However, the PCC between the two genes is 0.983, suggesting the two nodes are highly related. The reason is that the PCC method does not depend on the scale of expression values but detects the linear dependence of two genes. In contrast, our approach assumes that similar nodes have more shared elements in their context sets.}

\subsection{Results of network extraction}
\begin{figure}[ht]
\centering
\includegraphics[width=0.49\linewidth]{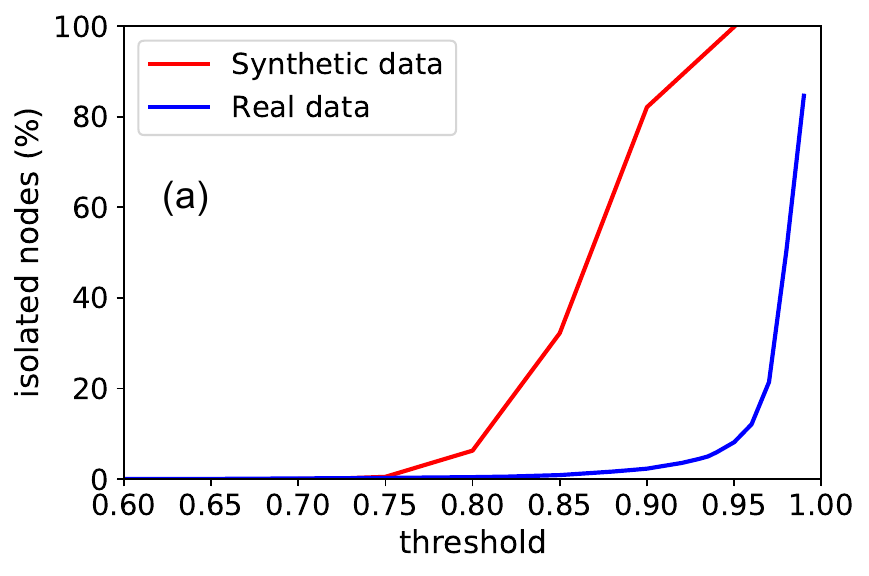}
\includegraphics[width=0.475\linewidth]{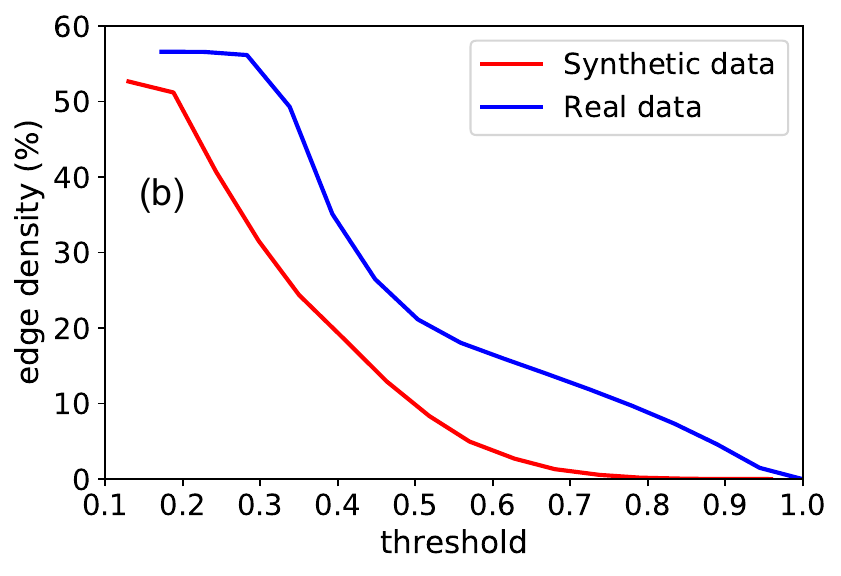}
\caption{\label{fig:fig8} Experimental results when different thresholds are applied. (a) shows the percentages of isolated nodes w.r.t. the thresholds. (b) shows the edge densities of networks when different thresholds are used.}
\end{figure}

\textcolor{blue}{Thresholding similarity value is the most straightforward and widely used approach in network construction. However, some nodes could be isolated from the network since these nodes may have relatively low similarities to all other nodes.} In Fig. \ref{fig:fig8}, we applied different thresholds to the cosine similarities computed from the node vectors trained with the synthetic data (Fig. \ref{fig:fig1}(a)) and the real data (Fig. \ref{fig:fig5}(a)). We observed that the percentage of isolated nodes increases rapidly when the thresholds are greater than 0.8 (synthetic data) and 0.95 (real data), respectively. In this paper, we define such threshold as the critical value. If we use a threshold smaller than the critical value, most nodes are connected to at least one other node. On the contrary, if the threshold is larger than the critical value, we possibly obtain a network with a large percentage of singleton nodes. 

\textcolor{blue}{We can force isolated nodes to connect with highly similar nodes in real applications. However, the neighbors selected through a single threshold are not affected by the distribution of similarity values. It is not rare that hubs are connected to many other nodes but with relatively low similarity values, while leaf nodes may connect to a small number of nodes with high similarities. That is, the distribution of similarity values is not considered in the selection of edges.} 

\begin{figure}[ht]
\centering
\includegraphics[width=0.475\linewidth]{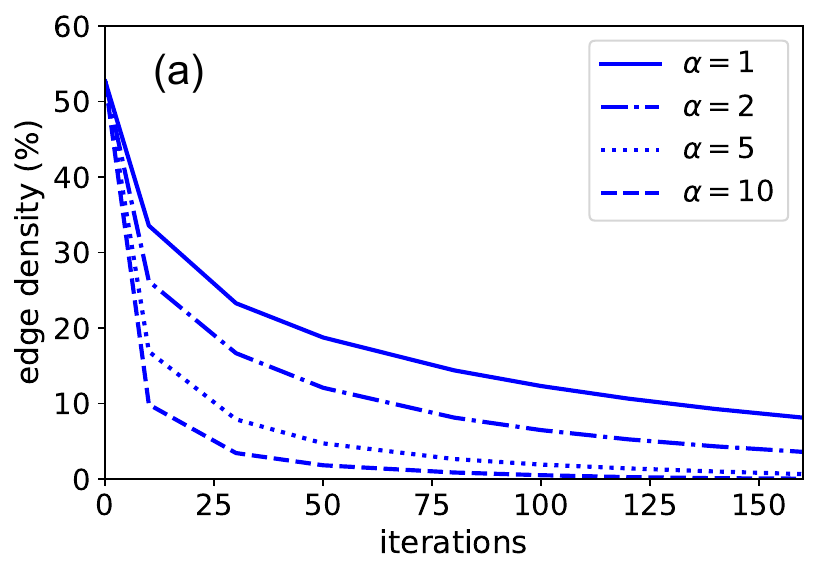}
\includegraphics[width=0.49\linewidth]{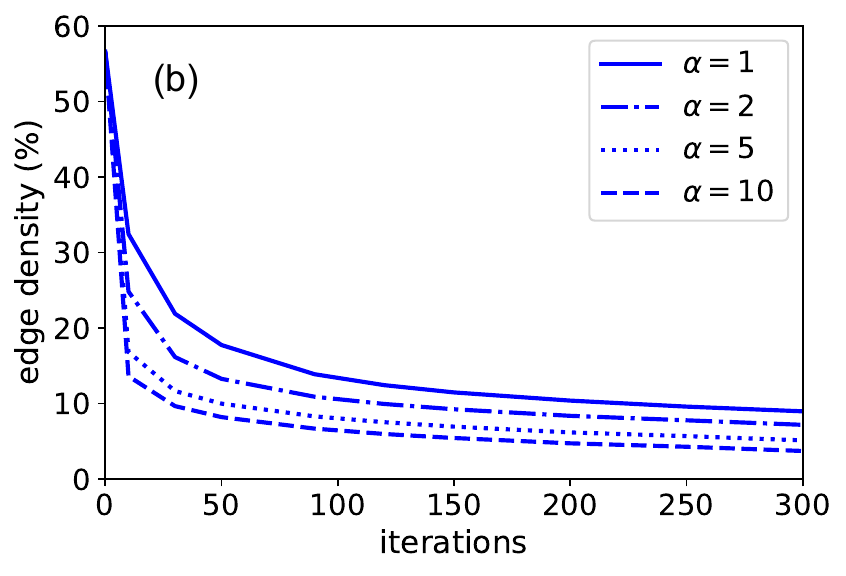}
\caption{\label{fig:fig9}Edge density analysis of the REM. (a) Synthetic data. (b) Real data.}
\end{figure}

\textcolor{blue}{The proposed REM will maintain every node connected to at least one other node since the "threshold" of each node is determined via the distribution of similarity values per Eq. \ref{eq:s16}.} In the experiments, we applied the REM to the two datasets used in Fig. \ref{fig:fig8}, and the results are shown in Fig. \ref{fig:fig9}. \textcolor{blue}{We observed that edge density decreases drastically in the first several iterations, and then the edge density decreases gradually. The reason is that the weakly connected edges are removed immediately from the network in the first several cycles. In contrast, the remaining edges have relatively high similarities, which are removed at a slower speed. In addition, the parameter $\alpha$ allows us to control the removal speed and edge density. Higher $\alpha$ removes weak links more efficiently, which aligns with our analysis in Sec. \ref{IIC}. In real applications, we can fix $\alpha$ and update Eq. \ref{eq:s14} to Eq. \ref{eq:s14} iteratively until we obtain a network with desired edge density.} 

\begin{table*}[ht]
\caption{\label{tab:tab3}The properties of networks constructed with the GTE and REM}
\begin{ruledtabular}
\begin{tabular}{ccccccccc}
&\multicolumn{2}{c}{$<$CT\footnote{CT denotes Critical threshold} (Syn.\footnote{Syn. denotes the synthetic data used in Fig. \ref{fig:fig8} and \ref{fig:fig9}})}&\multicolumn{2}{c}{$=$CT (Syn.)}&\multicolumn{2}{c}{$<$CT (Real. \footnote{Real. denotes the real data used in Fig. \ref{fig:fig8} and \ref{fig:fig8}})}&\multicolumn{2}{c}{$=$CT (Real)}\\
\textrm{Property}&
\textrm{GTE}&
\textrm{REM}&
\textrm{GTE}&
\textrm{REM}&
\textrm{GTE}&
\textrm{REM}&
\textrm{GTE}&
\textrm{REM}\\
\colrule
Threshold&0.7&-&0.8&-&0.92&-&0.95&-\\
Isolated nodes&0&0&287&0&372&0&849&0\\
Edge density&1.03\%&0.972\%&0.164\%&0.160\%&2.88\%&2.81\%&1.18\%&1.18\%\\
Ave. degree&51.5&48.6&8.2&8&300.8&293.2&123.2&123.4\\
Ave. clustering&0.46&0.41&0.38&0.32&0.66&0.55&0.59&0.37\\
\end{tabular}
\end{ruledtabular}
\end{table*}

\begin{figure}[ht]
\centering
\includegraphics[width=0.495\linewidth]{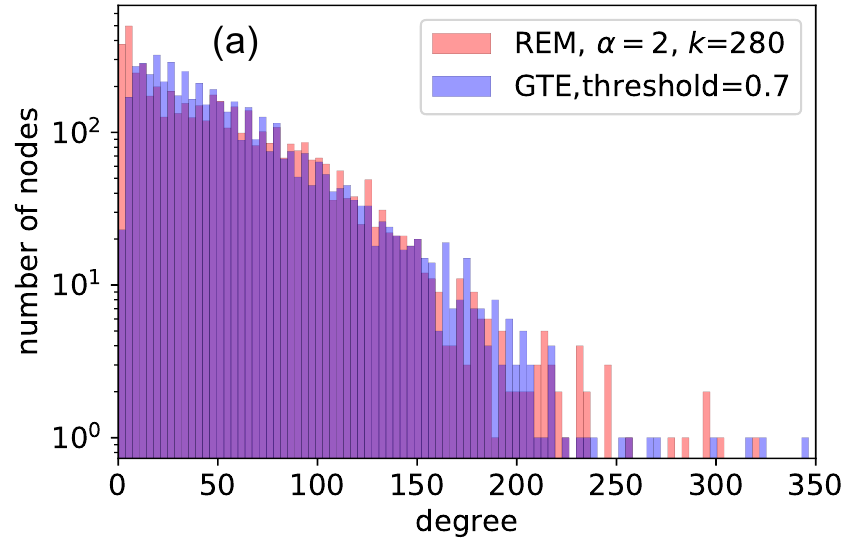}
\includegraphics[width=0.475\linewidth]{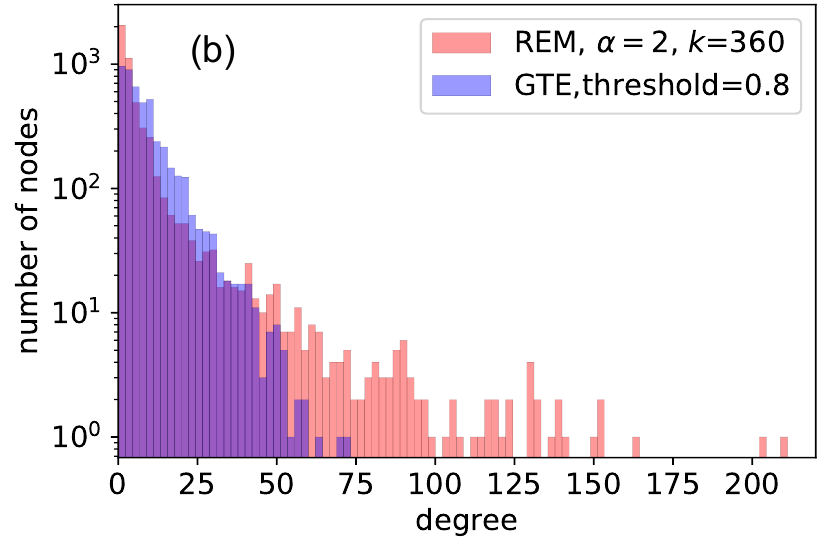}
\includegraphics[width=0.492\linewidth]{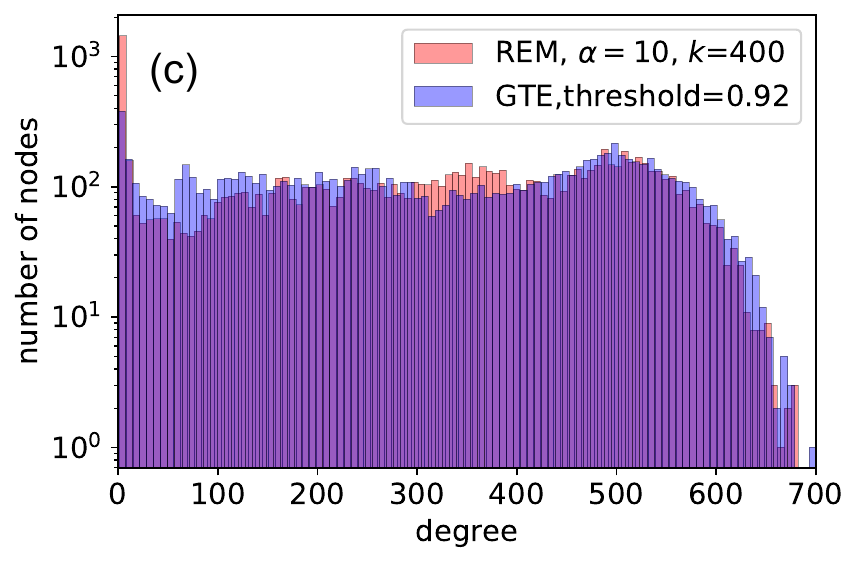}
\includegraphics[width=0.492\linewidth]{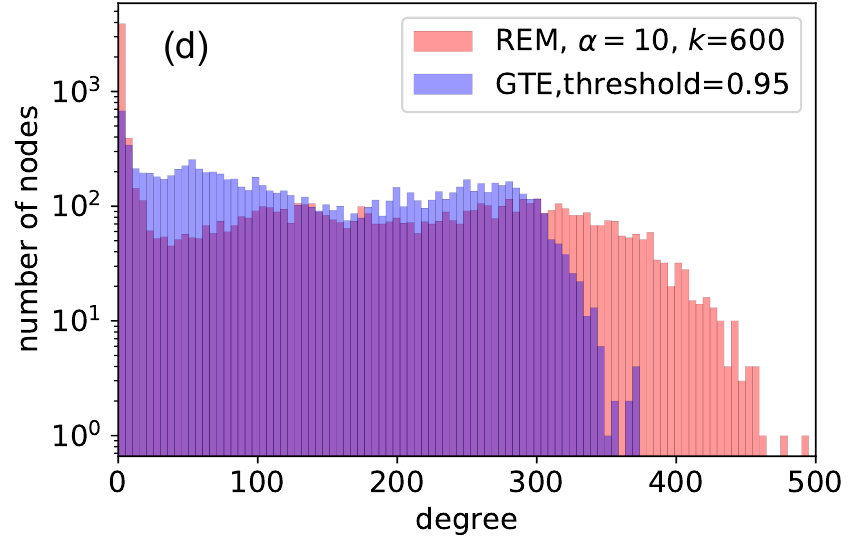}
\caption{\label{fig:fig10} \textcolor{blue}{The degree distributions of networks generated from the GTE and REM. (a) and (b) are the degree distributions of the networks constructed from synthetic data. (c) and (d) are the degree distributions of the networks constructed from real data. Panels (a) and (c) show the results of densely connected networks, while (b) and (d) are the results of sparsely connected networks.}}
\end{figure}

\textcolor{blue}{Furthermore, we generated four GTE-based networks with different thresholds for the datasets used in Fig. \ref{fig:fig8}.} The properties of the networks are shown in Table \ref{tab:tab3}. Specifically, The GTE networks are respectively generated with thresholds less and equal to the critical thresholds. In addition, we generated four REM networks, which have similar edge densities to their GTE counterparts. In Table \ref{tab:tab3}, we found that both GTE and REM return networks with similar average degrees when the edge densities are the same. However, the GTE networks always have a higher average clustering coefficient, suggesting that nodes in the GTE networks are more likely to cluster together. In Fig. \ref{fig:fig10}, we compared the degree distributions of the eight networks. We observed that the degree distributions of the GTE and REM networks almost overlap when the thresholds are less than the critical values (panels (a) and (c)). When the thresholds are at the critical values (panels (b) and (d)), some nodes in the REM networks still have high degrees, which are similar to the \emph{hubs} in many real networks. \textcolor{blue}{Besides, we observed that all four REM networks have many low-degree nodes, which account for the lower average clustering coefficients in Table \ref{tab:tab3}.}

\begin{figure}[ht]
\centering
\includegraphics[width=0.495\linewidth]{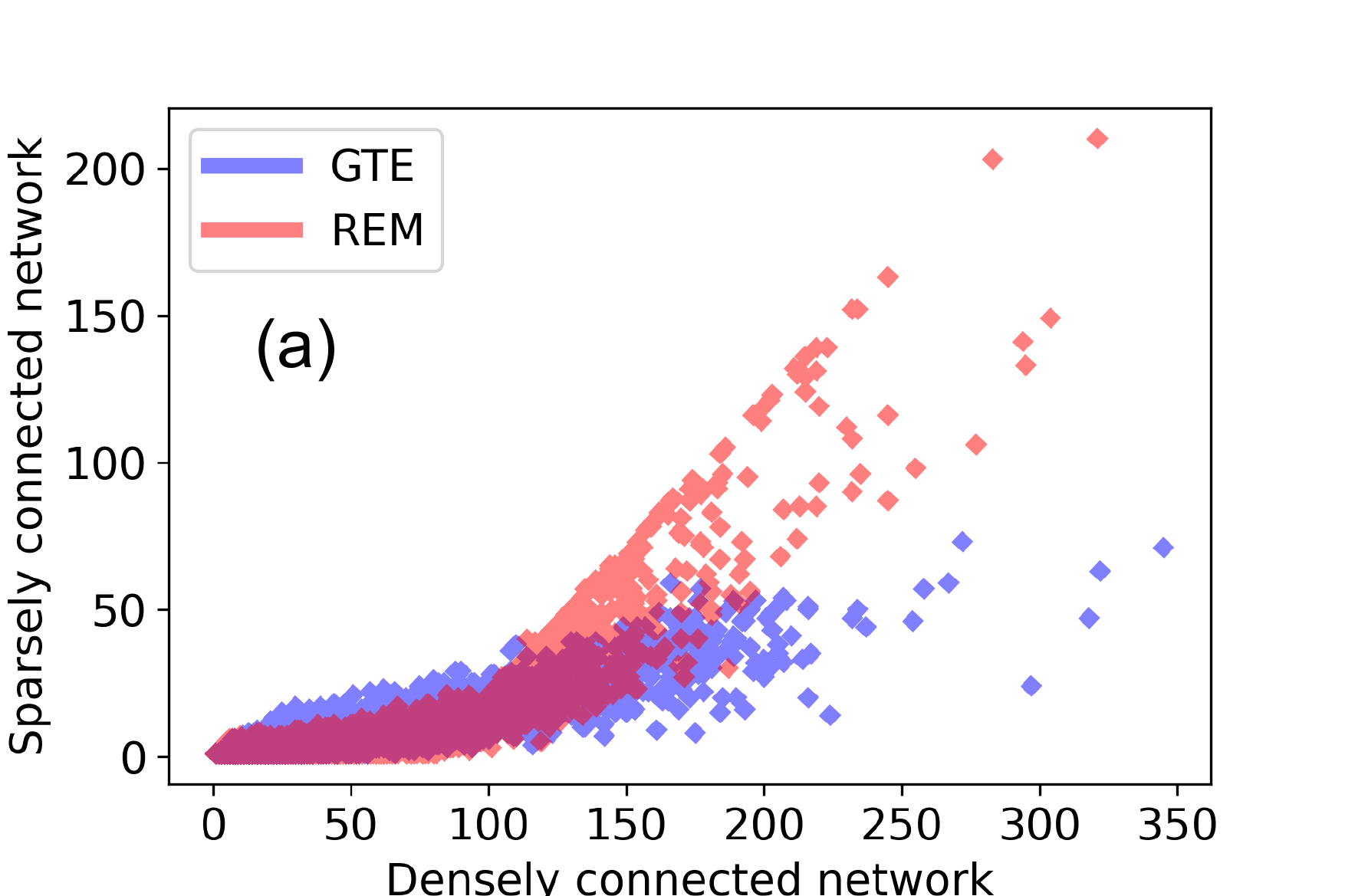}
\includegraphics[width=0.475\linewidth]{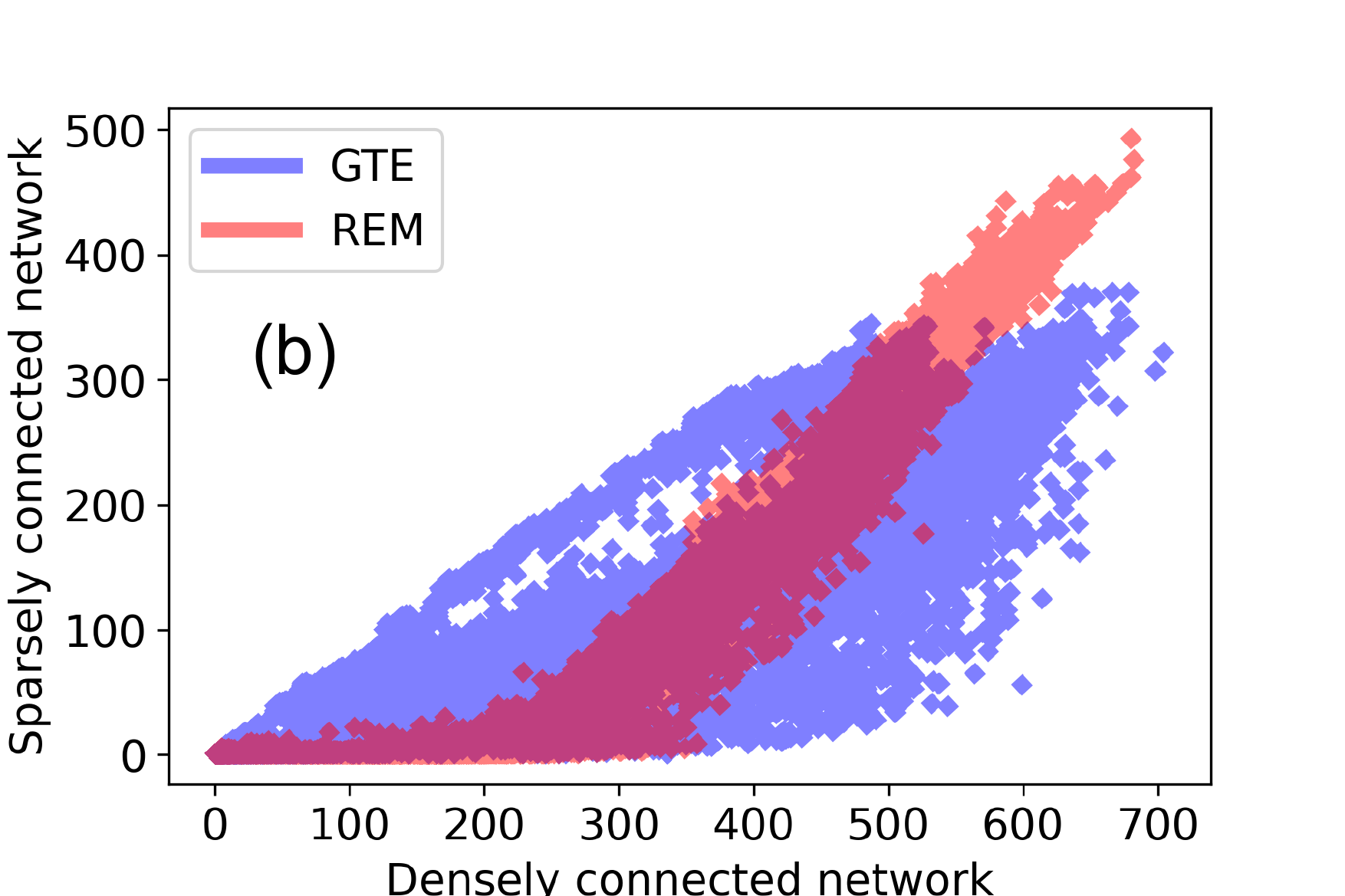}
\caption{\label{fig:fig11} \textcolor{blue}{The comparison of node degrees between densely and sparsely connected networks.}}
\end{figure}

\textcolor{blue}{Finally, we compared how edge density affects the roles of nodes. In Fig. \ref{fig:fig11}, each point represents a node. The horizontal coordinate represents the node’s degree in the densely connected network, while the vertical coordinate represents the node’s degree in the sparse network. We observed that the node degree of the GTE networks is remarkably affected by the threshold selection. The highest node degrees have dropped from 345 to 73 for the synthetic data and from 704 to 370 for the real data. In the REM network, the highest node degrees have dropped from 321 to 210 for the synthetic data and 682 to 493 for the real data. In Fig. \ref{fig:fig11}, the REM approach is more likely to remove edges from low-degree nodes. Edges from high-degree nodes are removed proportionally, which means the nodes’ roles are maintained and not significantly influenced by edge densities. On the other hand, the nodes’ degree in the resulting GTE networks is strongly related to the choice of edge density. In Fig. \ref{fig:fig11}(b), we can see that the points of the GTE networks are over-dispersed in the diagram.} 

\textcolor{blue}{More experimental results on real data are discussed in Appendix D.}

\section{Conclusion and future works}
\textcolor{blue}{This paper presents a neural network-based approach for learning node vectors from noisy node activity data. The primary advantage of the proposed method is that data are not required to follow any} specific distribution since we generate \emph{context sets} from raw data for each condition. \textcolor{blue}{The proposed approach is not constrained by missing values that ubiquitously exist in experimental results.} Inspired by the application of neural networks in natural language processing, we generate a corpus of node sequences to simulate sentences in documents. \textcolor{blue}{The corpus is trained by a neural network model, which produces node vectors and allows comparing and identifying nodes with synergistic roles. The experimental results show that the proposed approach is robust to the choice of parameters and missing values. In addition, we offer an alternative method to select edges for the underlying network. The REM method is based on the R\'enyi entropy and selects edges according to the distribution of similarity values. The proposed approach constructs networks without isolating nodes and can recover the roles of nodes.} 

In this work, we designed two experiments to test the proposed method. With both synthetic and real data, we showed that the proposed method could unveil the global and local structure of the nodal data even when 20\% values are randomly removed from the datasets. Furthermore, we tested the proposed \textcolor{blue}{entropy-based} network extraction method. \textcolor{blue}{We can obtain a network with desired edge density without isolated nodes by controlling the parameter $\alpha$ and the number of iterations.}

The experiments in this paper show promising results in detecting global and local structures from noisy nodal data. \textcolor{blue}{We expect the proposed data processing methodology to be used} in different areas, including biology and finance, especially where node activity data are measured with different techniques and missing values are present.

The code used for this work is available via a GitHub
613 repository https://github.com/BigBroKuang/embed-data-to-vector.

\section{Acknowledgements} 
This research is supported by the National Institutes of Health under Grant No. R01AI140760. \textcolor{blue}{The contents of this article are solely the responsibility of the authors and do not necessarily represent the official views of the funding agency.}

\appendix
\counterwithin{figure}{section}
\section{The skip-gram model} \label{apA}
\begin{figure}[ht]
\centering
\includegraphics[width=0.995\linewidth]{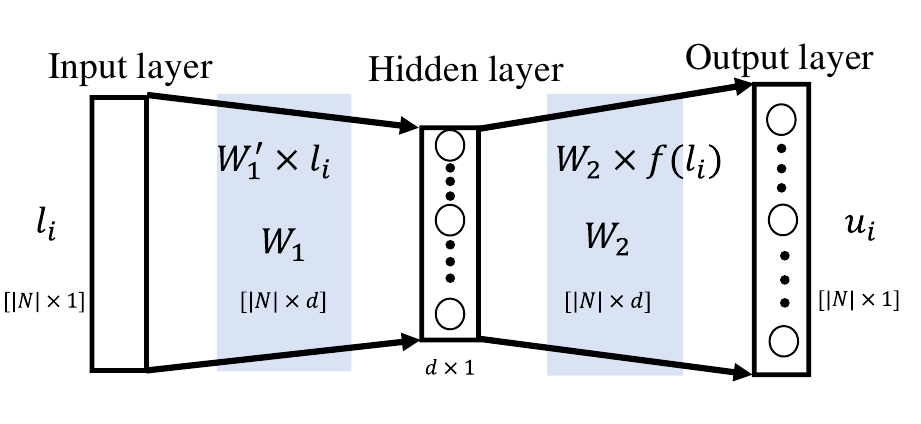}
\caption{\label{fig:figa1} The three-layer neural network model. Each input node is associated with an $N$ dimensional one-hot vector \cite{lan1}, which is mapped to the node vector $f(v_i)$ (the hidden layer) of dimension $d$ by matrix $W_1$. The hidden layer is mapped to the output vector by matrix $W_2$. The elements of $W_1$ and $W_2$ are initialized with random values, which are expected to be optimized by backpropagation \cite{lan2}.}
\end{figure}

The goal of generating random node sequences is to feed the corpus $T$ to neural networks to train node vectors. In this work, we adopt the simple three-layer skip-gram model \textcolor{blue}{as shown in Fig. \ref{fig:figa1}}. This neural network framework has three layers; input, hidden, and output layer \cite{lan1, lan2, lan3, barkan, bnb}. In this work, the goal is to find the $d$ dimensional vector \textcolor{blue}{for each of the $N$ nodes.}

In our assumption, nodes with similar values tend to appear in a similar context. Given a \textcolor{blue}{neighborhood} $H$ consisting of $2c$ nodes, \textcolor{blue}{we denote $P(v_x\ |\ H)$ as the conditional probability of node $v_x$ is neighboring to the $2c$ nodes in $H$.} Based on Bayes’ theorem, we have
\begin{align}
\label{eq:sr1}
P(v_x\ |\ H)=\frac{P(H\ |\ v_x)P(v_x)}{P(H)},
\end{align}
where $P(H)$ and $P(v_x)$ are respectively the probability of $H$ and $v_x$, and $P(H)$ and $P(v_x)$ can be regarded as constants. Then, we have
\begin{align}
\label{eq:sr2}
P(v_x\ |\ H)\propto P(H\ |\ v_x).
\end{align}

Now, we take one of the node sequences from the corpus. Let $l_i$ denote the $i$th node of the sequence, and $H=\{l_{i-c},…,l_{i-1},l_{i+1},…,l_{i+c}\}$. That is, we have an outcome $H$ given $l_i$. Since the goal is to determine $f(l_i)$, we replace $v_x$ in Eq. \ref{eq:sr2} with $f(l_i)$, and assume the $2c$ nodes are independent \cite{lan1}. \textcolor{blue}{We have}
\begin{align}
\label{eq:s10}
P(f(l_i)\ |\ H)\propto P(H\ |\ f(l_i)) =\nonumber\\
\prod_{-c \leq j \leq c, j \neq 0} P(l_{i+j}\ |\ f(l_i)),
\end{align}
where $P(l_{i+j}\ |\ f(l_i))$ is the occurring probability of node $l_{i+j}$ given the vector $f(l_i)$. To determine $f(l_i)$, we have the optimization problem after taking the log form of Eq. \ref{eq:s10}:
\begin{align}
\label{eq:s11}
E =-\min_f \sum_{-c \leq j \leq c, j \neq 0} \log P(l_{i+j}\ |\ f(l_i)).
\end{align}

In the model, the node vector $f(l_i)$ is projected to an $N$ dimensional output vector $u_i$ as shown in Fig. \ref{fig:figa1}. The $N$ dimensions of $u_i$ are associated to the $N$ nodes in the corpus. Then, we use the softmax function \cite{lan2,lan3, lecun} to map the entries of $u_i$ into probabilities, which all together give a probability distribution. For example, the probability of the $r$th entry of $u_i$ given $f(l_i)$ is
\begin{align}
\label{eq:s12}
P(v_r\ |\ f(l_i)) = \frac{\exp (u_i^r)}{\sum_{r^\prime \in N} \exp (u_i^{r^\prime})},
\end{align}
where $u_i^r$ is $r$th entry of $u_i$ and $P(v_r\ |\ f(l_i))$ is the probability of node $v_r$ to be the context of $l_i$. According to Eq. \ref{eq:s12}, nodes in $H$ have higher probabilities to be the context of node $l_i$. Combining Eq. \ref{eq:s11} and Eq. \ref{eq:s12}, we have the loss function \cite{dwang, lossfunc}
\begin{align}
\label{eq:s13}
E =  -\min_f \sum_{-c \leq j \leq c, j \neq 0} u_{i+j} +\log(\sum_{r^\prime \in N} \exp(u_i^{r^\prime})).
\end{align}
which is applied to every node in the sequence. Eq. \ref{eq:s13} is optimized by using the stochastic gradient descent approach \cite{vec1, lan2, lan3}, which backpropagates \cite{lecun} errors to update the elements of the matrices $W_1$ and $W_2$ in Fig. \ref{fig:fig2}.

The method we have introduced falls in the category of unsupervised learning, in which we learn node vectors from nodal data. The node vectors can be used to extract networks or detect nodes with similar properties.

\section{\textcolor{blue}{Synthetic datasets}} \label{apB}
\textcolor{blue}{The four synthetic datasets for the two case studies are generated according to Tables. \ref{tab:tabd1}, \ref{tab:tabd2}, \ref{tab:tabd3}, and \ref{tab:tabd4}.}
\begin{table}[!h]
\caption{\label{tab:tabd1}\textcolor{blue}{The synthetic dataset 1}}
\begin{ruledtabular}
\begin{tabular}{ccccccc}
&\multicolumn{6}{c}{Conditions}\\
group&$M_1$&$M_2$&$M_3$&$M_4$&$M_5$&$M_6$\\
\colrule
$G_1$& $A$&$B$&$C$&$D$&$E$&$R$\\
$G_2$&$	B$&$C$&$D$&$E$&$A$&$R$\\
$G_3$&$	C$&$D$&$E$&$A$&$B$&$R$\\
$G_4$&$D$&$E$&$A$&$B$&$C$&$R$\\
$G_5$&$	E$&$A$&$B$&$C$&$D$&$R$
\end{tabular}
\end{ruledtabular}
\end{table}

\begin{table}[!h]
\caption{\label{tab:tabd2}\textcolor{blue}{The synthetic dataset 2}}
\begin{ruledtabular}
\begin{tabular}{ccccccc}
&\multicolumn{6}{c}{Conditions}\\
group&$M_1$&$M_2$&$M_3$&$M_4$&$M_5$&$M_6$\\
\colrule
$G_1$&$\mathbf{A}$&$\mathbf{B}$&$C$&$D$&$E$&$R$\\
$G_2$&$\mathbf{A}$&$\mathbf{B}$&$D$&$E$&$D$&$R$\\
$G_3$&$B$&$C$&$E$&$A$&$C$&$R$\\
$G_4$&$C$&$D$&$A$&$B$&$B$&$R$\\
$G_5$&$D$&$E$&$B$&$C$&$A$&$R$
\end{tabular}
\end{ruledtabular}
\end{table}

\begin{table}[!h]
\caption{\label{tab:tabd3}\textcolor{blue}{The synthetic dataset 3}}
\begin{ruledtabular}
\begin{tabular}{ccccccc}
&\multicolumn{6}{c}{Conditions}\\
group&$M_1$&$M_2$&$M_3$&$M_4$&$M_5$&$M_6$\\
\colrule
$G_1$&$\mathbf{A}$&$\mathbf{B}$&$\mathbf{C}$&$D$&$E$&$R$\\
$G_2$&$\mathbf{A}$&$\mathbf{B}$&$\mathbf{C}$&$E$&$D$&$R$\\
$G_3$&$B$&$C$&$D$&$A$&$C$&$R$\\
$G_4$&$C$&$D$&$E$&$B$&$A$&$R$\\
$G_5$&$D$&$E$&$B$&$C$&$B$&$R$
\end{tabular}
\end{ruledtabular}
\end{table}

\begin{table}[!h]
\caption{\label{tab:tabd4}\textcolor{blue}{The synthetic dataset 4}}
\begin{ruledtabular}
\begin{tabular}{ccccccc}
&\multicolumn{6}{c}{Conditions}\\
group&$M_1$&$M_2$&$M_3$&$M_4$&$M_5$&$M_6$\\
\colrule
$G_1$&$\mathbf{A}$&$\mathbf{B}$&$\mathbf{C}$&$\mathbf{D}$&$E$&$R$\\
$G_2$&$\mathbf{A}$&$\mathbf{B}$&$\mathbf{C}$&$\mathbf{D}$&$D$&$R$\\
$G_3$&$B$&$C$&$D$&$E$&$C$&$R$\\
$G_4$&$C$&$D$&$E$&$A$&$B$&$R$\\
$G_5$&$D$&$E$&$B$&$C$&$A$&$R$
\end{tabular}
\end{ruledtabular}
\end{table}


\section{Parameter choice} \label{apC} 
\textcolor{blue}{We study the influence of $\rho$ on the relative distance between $G_1$ and $G_2$, and the results are shown in Table. \ref{tab:tab7}. We observe that the distance between the two communities slightly increases when we employ a low value of $\rho$ since a small $\rho$ encourages adding nodes that also exist in the context set of the previous node. As a result, far away nodes will become closer, reflected in the reduced distance between the two communities. However, the results are not significantly influenced by $\rho$ since the relative distances of two communities are maintained at the same level for the same data. Therefore, we recommend using $\rho=1$ in most cases.} 

\begin{table}[!h]
\caption{\label{tab:tab7}\textcolor{blue}{The relative distance between $G_1$ and $G_2$ w.r.t. $\rho$} }
\begin{ruledtabular}
\begin{tabular}{ccccccc}
\textrm{Data.}&
\textrm{1/10}&
\textrm{1/5}&
\textrm{1/3}&
\textrm{3}&
\textrm{5}&
\textrm{10}\\
\colrule
1&2.23&2.22&2.19&2.18&2.17&2.07\\
2&1.84&1.83&1.82&1.80&1.78&1.67\\
3&1.53&1.52&1.52&1.48&1.42&1.32\\
4&1.22&1.21&1.20&1.17&1.16&1.13\\
\end{tabular}
\end{ruledtabular}
\end{table}

\begin{table}[!h]
\caption{\label{tab:tab8}\textcolor{blue}{The prediction accuracy w.r.t. $\rho$}}
\begin{ruledtabular}
\begin{tabular}{ccccccc}
\textrm{Data.}&
\textrm{1/10}&
\textrm{1/5}&
\textrm{1/3}&
\textrm{3}&
\textrm{5}&
\textrm{10}\\
\colrule
1&98.24&98.24&98.72&98.80&98.90&98.82\\
2&98.26&97.22&98.68&98.00&97.30&98.12\\
3&97.52&97.48&97.72&98.22&97.58&97.32\\
4&78.14&78.58&78.62&79.72&79.32&79.78\\
\end{tabular}
\end{ruledtabular}
\end{table}

\section{\textcolor{blue}{Study the REM approach with AUC metrics on real data}} 
\textcolor{blue}{In this part, we use two real datasets with both network structure and node activity data to study the proposed approach. It is often hard to quantitatively determine the relationships between the network structure and node activity data because they describe the properties of nodes from different aspects. In the experiments, we learn node vectors from the node activity data, compute similarity and construct networks. The constructed network is compared to the network structures, and we use the AUC to evaluate our REM approach.}

\paragraph{\textcolor{blue}{The cora dataset.}} \textcolor{blue}{The cora dataset \cite{coradata} contains a sparse citation network with 2708 nodes and 5278 edges (the edge density is 0.144\%), where nodes represent publications and edges represent the citation relationships between the papers. Each node in the network is described by a 0/1-valued word vector, indicating the absence/presence of the corresponding word from a dictionary. The dictionary consists of 1433 unique words presented at least ten times in one of the 2708 publications.}

\paragraph{\textcolor{blue}{The pubmed dataset.}} \textcolor{blue}{The pubmed dataset \cite{pubmed} contains a sparse citation network with 19717 nodes and 44324 edges (the edge density is 0.0228\%), where nodes represent publications and edges represent the citation relationships between the papers. Each node in the network is described by a TF/IDF weighted word vector from a dictionary consisting of 500 words.}

\textcolor{blue}{The two networks represent citation relationships between the publications (nodes), while the node activity data are extracted from the content of each publication. We implement the proposed approach on these two real datasets to generate node vectors (128 dimensions). Then, we calculate the similarity for every node pair, and the performance of the approach is evaluated by comparing it to the true citation networks. The AUCs of the two datasets are respectively 0.81 and 0.73. Though the true relationship between network structure and node activity data is unknown, the results reveal that the node activity data are related to the network structure. Therefore, one of the advantages of our approach is that it allows us to compare two different types of data.}

\nocite{*}


\begin{thebibliography}{0}%
\makeatletter
\providecommand \@ifxundefined [1]{%
 \@ifx{#1\undefined}
}%
\providecommand \@ifnum [1]{%
 \ifnum #1\expandafter \@firstoftwo
 \else \expandafter \@secondoftwo
 \fi
}%
\providecommand \@ifx [1]{%
 \ifx #1\expandafter \@firstoftwo
 \else \expandafter \@secondoftwo
 \fi
}%
\providecommand \natexlab [1]{#1}%
\providecommand \enquote  [1]{``#1''}%
\providecommand \bibnamefont  [1]{#1}%
\providecommand \bibfnamefont [1]{#1}%
\providecommand \citenamefont [1]{#1}%
\providecommand \href@noop [0]{\@secondoftwo}%
\providecommand \href [0]{\begingroup \@sanitize@url \@href}%
\providecommand \@href[1]{\@@startlink{#1}\@@href}%
\providecommand \@@href[1]{\endgroup#1\@@endlink}%
\providecommand \@sanitize@url [0]{\catcode `\\12\catcode `\$12\catcode
  `\&12\catcode `\#12\catcode `\^12\catcode `\_12\catcode `\%12\relax}%
\providecommand \@@startlink[1]{}%
\providecommand \@@endlink[0]{}%
\providecommand \url  [0]{\begingroup\@sanitize@url \@url }%
\providecommand \@url [1]{\endgroup\@href {#1}{\urlprefix }}%
\providecommand \urlprefix  [0]{URL }%
\providecommand \Eprint [0]{\href }%
\providecommand \doibase [0]{https://doi.org/}%
\providecommand \selectlanguage [0]{\@gobble}%
\providecommand \bibinfo  [0]{\@secondoftwo}%
\providecommand \bibfield  [0]{\@secondoftwo}%
\providecommand \translation [1]{[#1]}%
\providecommand \BibitemOpen [0]{}%
\providecommand \bibitemStop [0]{}%
\providecommand \bibitemNoStop [0]{.\EOS\space}%
\providecommand \EOS [0]{\spacefactor3000\relax}%
\providecommand \BibitemShut  [1]{\csname bibitem#1\endcsname}%
\let\auto@bib@innerbib\@empty
\end{thebibliography}%


\begin{thebibliography}{4}
\bibitem{barabasi}
R. Pastor-Satorras, C. Castellano, P. Van Mieghem, and A.
Vespignani,
\href{https://doi.org/10.1103/RevModPhys.87.925}{Rev. Mod. Phys. 87, 925 (2015).}

\bibitem{barabasi1}
D.M. Gysi, Í. Do Valle, M. Zitnik, A. Ameli, X. Gan, O. Varol, S.D. Ghiassian, J.J. Patten, R.A. Davey, J. Loscalzo, and A.L. Barabási,
\href{https://doi.org/10.1073/pnas.2025581118}{Proc. Natl. Acad. Sci.  118, 19 (2021).}

\bibitem{bacco}
C. De Bacco, E. A. Power, D. B. Larremore, and C. Moore,
\href{https://doi.org/10.1103/PhysRevE.95.042317}{Phys. Rev. E 95, 042317 (2017).}

\bibitem{kuang}
J. Kuang and C. Scoglio,
\href{https://doi.org/10.1103/PhysRevE.104.024301}{Phys. Rev. E 104, 024301 (2021).}

\bibitem{newman1}
J. Young, G. T. Cantwell, and M.E.J. Newman,
\href{https://doi.org/10.1093/comnet/cnaa046}{Journal of Complex Networks 8, cnaa046 (2020).}
\bibitem{newman3}
M.E.J. Newman,
\href{https://doi.org/10.1103/PhysRevE.98.062321}{Phys. Rev. E 98, 062321 (2018).}
\bibitem{decelle}
A. Decelle, F. Krzakala, C. Moore, and L. Zdeborová,
\href{https://doi.org/10.1103/PhysRevE.84.066106}{Phys.
Rev. E 84, 066106 (2011).}

\bibitem{pardo}
R. Guimer\' a and M. Sales-Pardo,
\href{https://doi.org/10.1073/pnas.0908366106}{Proc. Natl. Acad. Sci. U.S.A. 106, 22073 (2009).}

\bibitem{peixoto1}
T.P. Peixoto,
\href{https://doi.org/10.1103/PhysRevE.97.012306}{Phys. Rev. E 97, 012306 (2018).}
\bibitem{peixoto2}
T.P. Peixoto,
\href{https://doi.org/10.1103/PhysRevX.8.041011}{Phys. Rev. X 8, 041011 (2018).}

\bibitem{van1}
B. Prasse and P. Van Mieghem,
\href{https://arxiv.org/abs/1807.08630}{arXiv:1807.08630}

\bibitem{peixoto3}
T. P. Peixoto, 
\href{https://arxiv.org/abs/1705.10225}{arXiv:1705.10225.}


\bibitem{newman4}
B. Karrer and M. E. Newman,
\href{https://doi.org/10.1103/PhysRevE.83.016107}{Phys. Rev. E 83, 016107 (2011)} 

\bibitem{peixoto4}
T.P. Peixoto,
\href{https://doi.org/10.1103/PhysRevLett.123.128301}{Phys. Rev. Lett. 123, 128301 (2018).}
\bibitem{timme1}
M. Timme,
\href{https://doi.org/10.1103/PhysRevLett.98.224101}{Phys. Rev. Lett. 98, 224101 (2007).}
\bibitem{timme2}
S. G. Shandilya and M. Timme,
\href{https://doi.org/10.1088/1367-2630/13/1/013004}{New J. Phys. 13, 013004 (2011).}
\bibitem{van2}
P. Van Mieghem and Q. Liu,
\href{https://doi.org/10.1103/PhysRevE.100.022317}{Phys. Rev. E 100, 022317 (2019).}
\bibitem{bianconi}
S. Boccaletti, G. Bianconi, R. Criado, CI. del Genio, J. Gómez-Gardeñes, and M. Romance,
\href{https://doi.org/10.1016/j.physrep.2014.07.001}{Phys Rep. 544, 1 (2014).}

\bibitem{newman2}
M.E.J. Newman,
\href{https://doi.org/10.1038/s41567-018-0076-1}{Nature Phys 14, 542 (2018).}

\bibitem{bob}
R.M. MacCallum, S.N. Redmond, and G.K. Christophides, 
\href{https://doi.org/10.1186/1471-2164-12-620}{BMC Genomics 14, 12, 620 (2011).}

\bibitem{horvath}
S. Horvath and J. Dong, 
\href{https://doi.org/10.1371/journal.pcbi.1000117}{PLoS computational biology 4, 8, e1000117 (2008).}

\bibitem{lynall}
M.E., Lynall,  D.S. Bassett,  R. Kerwin, P.J. McKenna, M. Kitzbichler, U. Muller, and E. Bullmore,  
\href{https://doi.org/10.1523/JNEUROSCI.0333-10.2010}{Journal of Neuroscience 30, 28, 9477 (2010).}

\bibitem{raimondo}
S. Raimondo, and M. De Domenico,
\href{https://doi.org/10.1523/JNEUROSCI.0333-10.2010}{Phys. Rev. E 103, 022311 (2021).}

\bibitem{eco}
G. Sugihara,R. May, H. Ye, C.H. Hsieh, E. Deyle, M. Fogarty, and S. Munch, 
\href{https://doi.org/10.1126/science.1227079}{science 338, 6106, 496 (2012).}

\bibitem{neuro}
E. Bullmore and O. Sporns,
\href{https://doi.org/10.1038/nrn2575}{Nat. Rev. Neurosci. 10, 186 (2009).}

\bibitem{neuro1}
B. Benigni, A. Ghavasieh, A. Corso, V. d’Andrea, and M. De Domenico,
\href{https://doi.org/10.1162/netn_a_00203}{Network Neuroscience 5, 3, 831 (2021).}

\bibitem{neuro2}
J. Schiefer, A. Niederbühl, V. Pernice, C. Lennartz, J. Hennig, P. LeVan, and S. Rotter, 2018. 
\href{https://doi.org/10.1371/journal.pcbi.1006056}{PLoS computational biology 14, 3, e1006056 (2018).}

\bibitem{mibrain}
J. Jeong, J.C. Gore,  and B.S. Peterson,
\href{https://doi.org/10.1016/S1388-2457(01)00513-2}{Clinical neurophysiology 112, 5, 827 (2001).}

\bibitem{fin}
A. Namaki, A.H. Shirazi, R. Raei, R. and G.R. Jafari,
\href{https://doi.org/10.1016/j.physa.2011.06.033}{Physica A 390, 21-22, 3835 (2011).}

\bibitem{climate}
K. Yamasaki, A. Gozolchiani, and S. Havlin,
\href{https://doi.org/10.1016/j.physa.2011.06.033}{Phys. Rev. Lett. 100, 228501 (2008).}

\bibitem{climate2}
\textcolor{blue}{J.F. Donges, Y. Zou, N. Marwan, and J. Kurths,}
\href{https://doi.org/10.1209/0295-5075/87/48007}{\textcolor{blue}{EPL 87, 48007, (2009).}}

\bibitem{newmannc}
J.G. Young, A. Kirkley, and  M.E.J. Newman,
\href{https://doi.org/10.1103/PhysRevE.105.014312}{Phys. Rev. E 105, 014312 (2022).}

\bibitem{newmanpre1}
M.E. Newman,
\href{https://doi.org/10.1103/PhysRevE.98.062321}{Phys. Rev. E 98, 062321 (2018).}

\bibitem{barabasi2}
I.A. Kovács, D.L. Barabási, and A.L. Barabási,
\href{https://doi.org/10.1073/pnas.2009093117}{Proc. Natl. Acad. Sci.  117, 33570 (2020).}
\bibitem{horvath1}
M.C. Oldham, S. Horvath, and D.H. Geschwind,
\href{https://doi.org/10.1073/pnas.0605938103}{Proc. Natl. Acad. Sci. 103, 17973 (2006).}

\bibitem{horvath2}
B. Zhang, and S. Horvath, 
\href{https://doi.org/10.2202/1544-6115.1128}{Statistical applications in genetics and molecular biology, 4, 1,(2005)}

\bibitem{horvath3}
L. Song, P. Langfelder, and  S. Horvath,
\href{https://doi.org/10.1186/1471-2105-13-328}{BMC bioinformatics, 13, 1 (2012)}
\bibitem{info}
A.A. Margolin, I. Nemenman, K. Basso, C. Wiggins,  G. Stolovitzky, R. Dalla Favera, and A. Califano, 
\href{https://doi.org/10.1186/1471-2105-7-S1-S7}{ BMC Bioinformatics 7, S7 (2006).}
\bibitem{cancer}
 R. Ramirez, Y.C. Chiu, A. Hererra, M. Mostavi,J. Ramirez,Y. Chen, Y. Huang,  and  Y.F. Jin, 
 \href{https://doi.org/10.3389/fphy.2020.00203}{Frontiers in physics 8, 203 (2020).}

\bibitem{reverter}
A. Reverter and E.K. Chan,
\href{https://doi.org/10.1093/bioinformatics/btn482}{Bioinformatics 24, 21, 2491 (2008).}

\bibitem{misdata}
S. García, J, Luengo, and F. Herrera, 
\href{https://doi.org/10.1007/978-3-319-10247-4_4}{Intelligent Systems Reference Library, 72. (2015).}

\bibitem{kuangano}
 \textcolor{blue}{J. Kuang, N. Buchon, K. Michel, and C. Scoglio,}
\href{https://doi.org/10.1186/s12859-022-04697-9}{\textcolor{blue}{BMC Bioinformatics 23, 170 (2022).}}

\bibitem{occam}
 \textcolor{blue}{B. Dresp-Langley, O.K. Ekseth, J. Fesl, S. Gohshi, M. Kurz, and H.-W. Sehring,}
\href{https://doi.org/10.3390/app9153065}{\textcolor{blue}{Appl. Sci., 9, 3065. (2019).}}
 
\bibitem{dam}
\textcolor{blue}{S. Van Dam, U. Vosa, A. van der Graaf, L. Franke, and J.P. de Magalhaes,} \href{https://doi.org/10.1093/bib/bbw139}{\textcolor{blue}{Briefings in bioinformatics, 19, 4, 575 (2018).}}

\bibitem{vec1}
A. Grover and J. Leskovec,
\href{https://doi.org/10.1145/2939672.2939754}{In Proceedings of the 22nd ACM SIGKDD International Conference on Knowledge Discovery and Data Mining (KDD '16) 855, (2016).}

\bibitem{vec2}
B. Perozzi, R. Al-Rfou, and S. Skiena,
\href{https://doi.org/10.1145/2623330.2623732}{In Proceedings of the 20th ACM SIGKDD international conference on Knowledge discovery and data mining (KDD '14) 701, (2014).}

\bibitem{vec3}
J. Tang, M. Qu, M. Wang, M. Zhang, J. Yan, and Q. Mei,
\href{https://doi.org/10.1145/2736277.2741093}{In Proceedings of the 24th International Conference on World Wide Web (WWW '15) 1067, (2015).}

\bibitem{lan1}
M.E. Peters, M. Neumann, M. Iyyer, M. Gardner, C. Clark, K. Lee, and L. Zettlemoyer,
\href{http://arxiv.org/abs/1802.05365}{ArXiv:1802.05365}

\bibitem{lan2}
R, Xin, 
\href{http://arxiv.org/abs/1411.2738}{ArXiv:1411.2738}

\bibitem{lan3}
T. Mikolov, I. Sutskever, K. Chen, G. Corrado, and J. Dean,
\href{https://arxiv.org/abs/1310.4546}{ArXiv:1310.4546}

\bibitem{lee}
\textcolor{blue}{M.J. Lee, E. Lee, B. Lee, H. Jeong, D.S. Lee, and S.H. Lee,}
\href{https://doi.org/10.1103/PhysRevResearch.3.043136}{\textcolor{blue}{Phys. Rev. Research 3, 043136 (2021)}}

\bibitem{entropy2}
M. De Domenico, V. Nicosia, A. Arenas, and V. Latora,
\href{https://doi.org/10.1038/ncomms7864}{Nat. Commun. 6, 6864 (2015)}

\bibitem{mark}
A. Shamshad, M.A. Bawadi, W.M.A. Wan Hussin, T.A. Majid, and S.A.M.Sanusi,
\href{https://doi.org/10.1016/j.energy.2004.05.026}{Energy 30, 5, 693 (2005) }

\bibitem{barkan}
\textcolor{blue}{O. Barkan and N. Koenigstein,}
\href{10.1109/MLSP.2016.7738886}{\textcolor{blue}{2016 IEEE 26th International Workshop on Machine Learning for Signal Processing (MLSP), (2016) }}

\bibitem{bnb}
\textcolor{blue}{M. Grbovic and H. Cheng,}
\href{https://doi.org/10.1145/3219819.3219885}{\textcolor{blue}{In Proceedings of the 24th ACM SIGKDD International Conference on Knowledge Discovery and Data Mining (KDD '18),(2018) }}

\bibitem{hill}
\textcolor{blue}{M.O. Hill,}
\href{https://doi.org/10.2307/1934352}{\textcolor{blue}{Ecology 54, 427 (1973).}}

\bibitem{entropy3}
K. Zyczkowski,
\href{https://doi.org/10.1023/A:1025128024427}{\textcolor{blue}{Syst. Inf. Dyn 10, 297 (2003).}}

\bibitem{hillskew1}
\textcolor{blue}{R.J. Hill,}
\href{https://doi.org/10.1016/j.jeconom.2004.08.018}{\textcolor{blue}{Journal of Econometrics 130, 1, 25, (2006).}}
\bibitem{hillskew2}
\textcolor{blue}{A. Chao, N.J. Gotelli, T.C. Hsieh, E.L. Sander, K.H. Ma, R.K. Colwell, and A.M. Ellison,}
\href{ https://doi.org/10.1890/13-0133.1}{\textcolor{blue}{Ecological Monographs, 84, 45, (2014).}}

\bibitem{entropy4}
L. Jost,
\href{https://doi.org/10.1111/j.2006.0030-1299.14714.x}{Oikos 113, 363 (2006).}

\bibitem{pca}
S. Wold, K. Esbensen,  and P. Geladi, 
\href{https://doi.org/10.1016/0169-7439(87)80084-9}{Chemometrics and intelligent laboratory systems 2, 1-3, 37 (1987).}
\bibitem{filz}
P. Filzmoser, K. Hron, and C. Reimann,
\href{https://doi.org/10.1002/env.966}{The Official Journal of the International Environmetrics Society 20, 6, 621 (2009).}

\bibitem{kmean}
\textcolor{blue}{A. Likas, N. Vlassis, and J. J. Verbeekb,}
\href{https://doi.org/10.1016/S0031-3203(02)00060-2}{\textcolor{blue}{Pattern Recognition 36, 2, 451, (2003).}}

\bibitem{bon1}
K. Baltakys, J. Kanniainen, and F. Emmert-Streib,  
\href{https://doi.org/10.1038/s41598-018-26575-2}{Sci Rep 8, 8198 (2018).}

\bibitem{ball}
B. Ball, B. Karrer, and M.E. Newman,
\href{https://doi.org/10.1103/PhysRevE.84.036103}{Phys. Rev. E 84, 036103 (2011).}

\bibitem{wang}
 M.H. Wang, O. Marinotti, A. Vardo-Zalik, R. Boparai, and G. Yan,
\href{https://doi.org/10.1371/journal.pone.0026011}{PLoS one, 6, e26011 (2011).}

\bibitem{rogers}
D.W. Rogers, M.M. Whitten, J. Thailayil, J. Soichot, E.A. Levashina, and F. Catteruccia,
\href{https://doi.org/10.1073/pnas.0809723105}{Proc. Natl. Acad. Sci. 105, 19390 (2008).}

\bibitem{maaten}
L. Van Der Maaten, 
\href{https://proceedings.mlr.press/v5/maaten09a.html}{In Artificial Intelligence and Statistics, 384 (2009).}

\bibitem{lecun}
Y. LeCun, Y. Bengio, and G. Hinton, 
\href{https://doi.org/10.1038/nature14539}{Nature 521, 436 (2015) }

\bibitem{dwang}
\textcolor{blue}{D. Wang,}
\href{https://doi.org/10.1609/aimag.v22i2.1565}{\textcolor{blue}{AI Magazine 22, 2, 10, (2001).}}

\bibitem{lossfunc}
\textcolor{blue}{Q. Wang, Y. Ma, K. Zhao, and Y. Tian,}
\href{https://doi.org/10.1007/s40745-020-00253-5}{\textcolor{blue}{Annals of Data Science, 9, 187 (2022) }}

\bibitem{ruan}
\textcolor{blue}{J. Ruan, A.K. Dean,  and  W. Zhang,}
\href{https://doi.org/10.1186/1752-0509-4-8}{\textcolor{blue}{BMC Syst. Biol. 4, 8 (2010).}}

\bibitem{coradata}
\href{https://relational.fit.cvut.cz/dataset/CORA}{\textcolor{blue}{The cora dataset}}

\bibitem{pubmed}
\href{https://relational.fit.cvut.cz/dataset/PubMed_Diabetes}{\textcolor{blue}{The pubmed dataset}}



\end{thebibliography}

\end{document}